\documentclass[aps,prb,twocolumn,amsmath,amssymb,nofootinbib,superscriptaddress,floatfix]{revtex4-1}
\usepackage{amssymb}

\usepackage{amssymb}
\usepackage{graphicx}
\usepackage{mathtools}

\def\beq{\begin{equation}}
\def\eeq{\end{equation}}

\begin{document}

\title{Relationship between
Symmetry Protected Topological Phases and Boundary Conformal Field Theories via the Entanglement Spectrum}

\author{Gil Young Cho}
\affiliation{Department of Physics, Korea Advanced Institute of Science and Technology, Daejeon 305-701, Korea}
\author{Ken Shiozaki} 
\affiliation{Department of Physics and Institute for Condensed Matter Theory, University of Illinois, 1110 W. Green St., Urbana, Illinois 61801-3080, USA}
\author{Shinsei Ryu}
\affiliation{Department of Physics and Institute for Condensed Matter Theory, University of Illinois, 1110 W. Green St., Urbana, Illinois 61801-3080, USA}
\author{Andreas W.W. Ludwig} 
\affiliation{Physics Department, University of California, Santa Barbara, CA 93106, USA}

\date{\today}

\begin{abstract}
Quantum phase transitions out of a symmetry-protected topological (SPT) phase in (1+1) dimensions into
an adjacent,  topologically distinct 
SPT phase protected by the same symmetry or a  trivial gapped phase, are typically 
described by a conformal field theory (CFT). 
At the same time, the low-lying entanglement spectrum of a  gapped phase close
to such  a  quantum critical point is known(Cho et al., arXiv:1603.04016), very generally,  to be universal and
described by  (gapless)  boundary conformal field theory.
Using this connection we show that symmetry properties of 
the boundary conditions in boundary CFT  can be used to characterize the  symmetry-protected degeneracies of the entanglement
spectrum, a hallmark of non-trivial symmetry-protected topological phases. Specifically, we show that the relevant boundary CFT
is the orbifold of the quantum critical point with respect to the symmetry group defining the SPT, and that the boundary states of this orbifold
carry a  quantum anomaly that determines the topological class of the SPT.
We illustrate this connection
using  various characteristic  examples such as the time-reversal breaking
``Kitaev chain'' superconductor (symmetry class D), the Haldane phase, and the $\mathbb{Z}_8$ classification of 
interacting topological superconductors  in symmetry class BDI in (1+1) dimensions.

\end{abstract}


\maketitle

\tableofcontents
\section{Introduction}

The recent progress in our understanding of phases of matter
has revealed that there are plenty of phases that go beyond Landau's  symmetry breaking paradigm.  
\cite{Wen2004Book} 
Having various quantum disordered phases,
which are not characterized by spontaneous symmetry breaking, 
we can ask if all these phases are (topologically) equivalent or not. 
At least for gapped phases of matter, which are our focus in this paper, 
the (partial) answer to this question is known. 
There are at least three broad classes of quantum disordered phases: 
(i)  topologically trivial phases, 
(ii) phases
with intrinsic topological order,
including symmetry-enriched topological phases, the fractional quantum Hall effect,
and (iii)
symmetry protected topological (SPT) phases, 
including electronic topological band insulators.
The literature on these classes of phases of matter is by now too 
exhaustive to mention here, but for example, see
Refs.\ \onlinecite{Wen2004Book, KaneRev, QiRev, Turner2013, SenthilReview, Chiu2015} 
for reviews,
and Refs.\ \onlinecite{
Kitaev2006,
Mesaros2013,
Lu2013classification,
Essin2013classifying,
Cho2012gapless}
and
\onlinecite{
Classification2, Kitaev2009, 
Ryu2012, Qi2012, CWang2014, Fidkowski2013, Metlitski2014, 
Vishwanath2013, Geraedts2014, Lu2012a, Cho2014, Cappelli2013,
Chen2013, Kapustin2014a, Kapustin2014b, Kapustin2014c,
Barkeshli2014,Teo2015},
for recent studies on symmetry enriched topological phases and SPT phases, respectively.

In this paper, we will establish a link between 
(1+1)d SPT phases which are gapped, and boundary conformal field theories (CFTs) which are gapless.
In particular, we will  associate specific types of boundary conditions
in boundary CFTs (BCFTs) to an SPT phase.

A motivation to connect gapped SPT phases to CFT (or BCFT), which describes gapless critical points or critical
phases, comes naturally from the following observation.
By definition, 
distinct SPT phases cannot be adiabatically deformed into each other
while preserving the symmetries that define the SPT phase,
without going through a quantum critical point at which the gap closes:
In other words, distinct SPT phases are  separated by a
quantum critical point which is typically a CFT.
Thus  in the phase diagram,  a given SPT phase is typically 
in proximity of a CFT. 
(The SPT belongs to the ``theory space'' of quantum field theories that
can be reached  from the CFT by applying perturbations relevant in the renormalization group sense.)
One may then wonder to which extent a given CFT describing such a quantum critical point
knows about SPT phases  which are located just in its immediate neighborhood.
Since an arbitrarily   small gap is enough to define a topological phase, 
the question which relevant operator (``massive deformation'')
of a given CFT  
gives rise to a specific topological or trivial phase
in its vicinity 
can be deduced solely from data contained in the CFT.

In this paper, we associate a particular BCFT  with  a given  (1+1) dimensional SPT phase by using  a number
of different arguments.
One of our main arguments,
which we believe to be the most fundamental and universal, 
uses the entanglement spectrum.
The entanglement spectrum 
has generally been proven to be a useful tool to study SPT phases.
\cite{Ryu2006, pollmann2012symmetry}
In particular, it has been previously claimed and proven, using  matrix product states (MPSs), 
that the entanglement spectrum of the ground state of a (1+1) dimensional  SPT phase is degenerate, 
and  that the degeneracy of the entanglement spectrum is protected by the symmetries which define the SPT phase
(``symmetry protected degeneracy'').
In this paper, by establishing a connection to BCFTs,  
we will develop an analytical understanding of the entanglement spectrum of 
SPT phases  near their proximate quantum critical points, which are described by a CFT.
We can then use the knowledge of the corresponding BCFTs to study SPT phases.

In another argument, 
we try to detect non-trivial properties of a given SPT phase 
by first attaching an ideal ``lead'' (a gapless quantum field theory) to the SPT phase (see FIG. \ref{fig:S-matrix}).
We then ``shoot''  quasiparticles (e.g., electrons) from the ``lead''  into the SPT phase and measure their scattering off 
from the SPT phase to learn something about the SPT phase.
Such an approach has been applied to non-interacting fermionic SPT phases in all dimensions 
\cite{Akhmerov2011}
and has proven to be quite powerful. 
E.g., from the properties of the scattering matrix, one can obtain the 10-fold classification 
of topological insulators and superconductors.\cite{Classification2,TenFoldWayNJPhys2010,Kitaev2009}
In the present  paper, we generalize this approach to (1+1)-d SPT phases with interactions,
by using BCFTs. 

As an application and illustration  of our framework,  
we will discuss archetypical topological states in one spatial dimension, 
such as the time-reversal breaking
topological superconductor in symmetry  class D (a fermionic SPT phase)\cite{kitaev2001unpaired}, 
and the Haldane chain (a bosonic SPT phase).
We will also apply our framework  to 
topological superconductors in symmetry class BDI  in (1+1) dimensions. 
For this system, 
Fidkowski and Kitaev 
\cite{Fidkowski2010, fidkowski2010entanglement}
found a ``counter example'' of the non-interacting classification
of topological insulators/superconductors.
While at the non-interactive level, 1d topological superconductors
in symmetry class BDI
are classified by an integer topological, invariant, 
Fidkowski and Kitaev found that,
with interaction, the $\mathbb{Z}$ classification reduces to 
the smaller  $\mathbb{Z}_8$ classification. 
It would be quite interesting to understand 
in further detail
how the non-interacting classification reduces to this smaller classification  in
the presence of interactions. 
By linking SPT phases to BCFTs,
we deepen our understanding of this phenomenon.  

For loosely related works, see, for example,  Ref.\ \onlinecite{DeGottardi2014}, 
and Ref.\ \onlinecite{Ohmori2015}. (The latter work studied the 
role of boundaries in the entanglement spectrum in (1+1)-dimensional  (gapless)
CFTs, as opposed to the  gapped (1+1)-dimensional  SPT phases discussed in the present work.)
In Ref.\ \onlinecite{2015JHEP...05..152M},
the relationship between gapped phases in (1+1) dimensions and boundary states in boundary CFTs
was discussed in the context of 
the (continuous) MERA tensor network representation of
quantum ground states and their holographic duality.

The rest of the paper is organized as follows.  

In Sec.\ \ref{BCFTs and SPTs}, 
we provide various setups allowing us  to make a connection between SPT phases and BCFTs.
In particular, we consider 
gapless CFTs which are in contact with gapped SPT phases, 
and we discuss
the entanglement spectrum of gapped (1+1) dimensional SPT phases close to a quantum critical point.
(See also Ref. [\onlinecite{ChoLudwigRyuarXiv2016}].) 

In Sec.\ 
\ref{LabelSection-SymmetryProtectedDegeneracyEntanglementSpectrumGeneralConsiderations}
we discuss the problem of identifying a proper boundary state of the CFT for a given SPT phase. 
This is also related to the question as to how 
we describe, in the language of CFT,  the symmetry-protected degeneracy of the entanglement 
spectrum, a hallmark of (1+1)-dimensional SPT phases.
In order to achieve this goal, we will propose to use  boundary states 
of an orbifold CFT,
which is obtained from the original CFT by orbifolding it by the symmetry group defining the SPT.

This methodology is demonstrated in the two simplest examples of 
(1+1) dimensional SPT phases, namely, 
the Kitaev chain (Sec. \ref{The Kitaev chain}),
and the Haldane chain (Sec.\ \ref{The Haldane phase and the compact boson theory}). 
We will also discuss
the $\mathbb{Z}_8$ classification of Fidkowski-Kitaev in the class BDI Majorana chain
in Sec.\ \ref{BDI topological superconductors in (1+1)d}. 
The symmetry group of this system involves time-reversal, which 
needs to be treated somewhat 
differently from unitary on-site symmetries. 
We conclude in Sec.\ \ref{Discussion}.

\section{Boundary Conformal Field Theories  (BCFTs) and Symmetry Protected Topological Phases (SPTs)} 
\label{BCFTs and SPTs}

In this section, we give an overview of a set of arguments
which support the advocated relation between BCFTs and SPTs:
(A) The Jackiw-Rebbi domain wall;
(B) 
the scattering from SPT phases;
and 
(C)
the entanglement spectrum. 

\subsection{Interface between trivial and topological phases}
\label{LabelSectionExpandedDomainWallPicture}

\begin{figure}
\begin{center}
\includegraphics[width=\columnwidth]{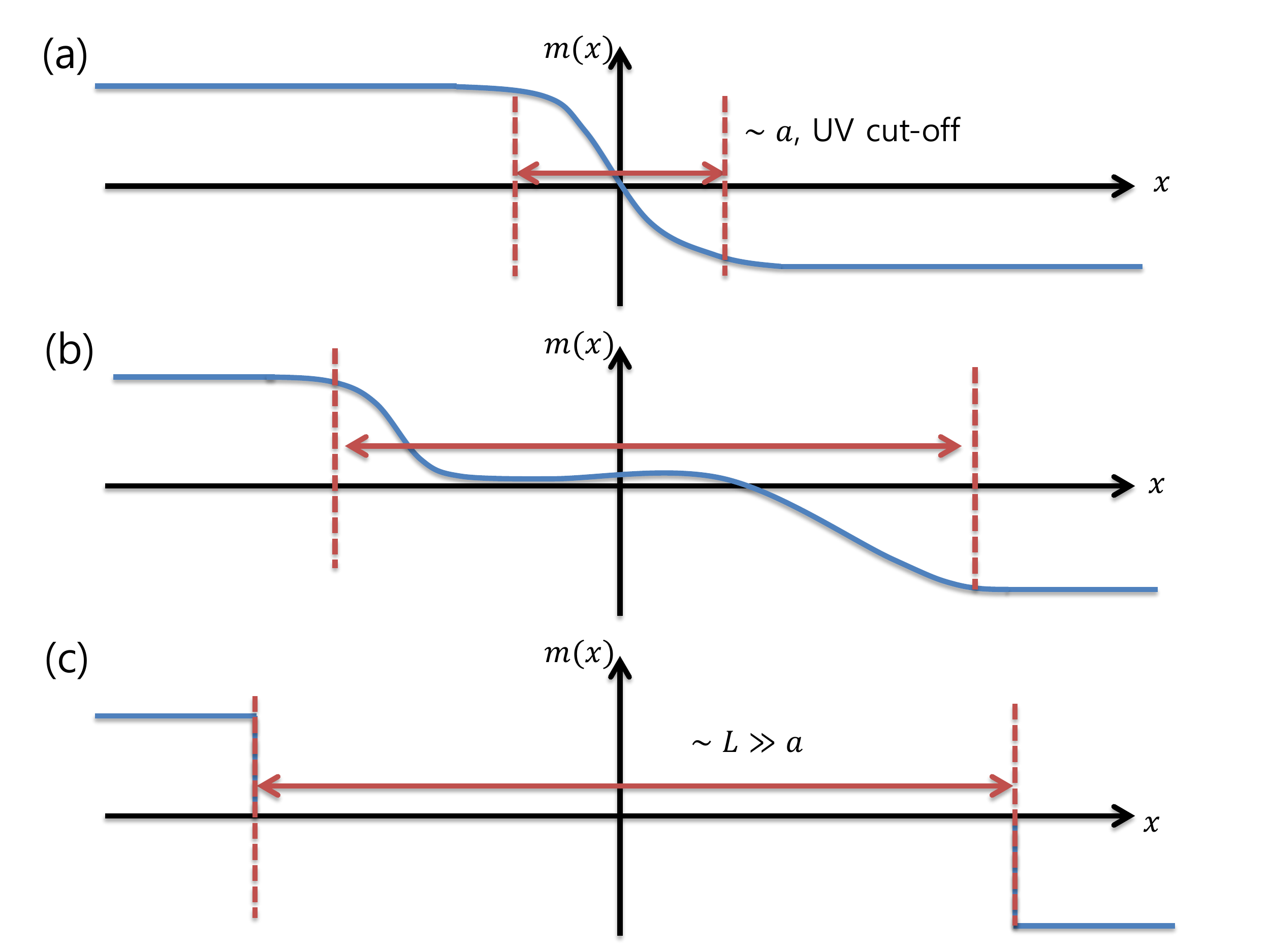}
\caption{
Deformation of the domain wall. 
(a) A domain wall with the size $a$, which is of the lattice scale. 
SPT phases will localize a zero mode at the domain wall. 
(b) The domain wall can smoothly be deformed to a bigger spatial region. 
In this manipulation of the domain wall, the topological zero mode cannot be removed. 
(c) When we push the domain wall to $L\gg a$, we effectively find a critical mode localized at the length scale $L$. Even in this limit, the topological zero mode will be superposed with the critical mode whose level spacing will be determined by the non-topological scale $\sim 1/L$. This picture suggests that the boundary zero mode of the SPT phases can be thought as the critical mode localized at the UV scale `$a$' as mentioned in the main text. 
 \label{Fig1}
}
\end{center}
\end{figure}

A principle that underlies all non-interacting topological phases of Fermions
can be well illustrated by the Jackiw-Rebbi domain wall 
(and its analogue in different dimensions and in different symmetry classes);
let us consider a massive Majorana fermion system
described by
the action $S=S_*+S_I$
\begin{align}
S_* 
&=
\frac{1}{4\pi}
\int dt dx 
\big[
\psi_L {i} (\partial_t -v \partial_x) \psi_L
+
\psi_R {i} (\partial_t + v \partial_x) \psi_R
\big], 
\nonumber \\
S_I
&=
- {i} m \int dt dx\,  
\psi_L\psi_R, 
\label{Kitaev chain}
\end{align}
where $\psi_L$ ($\psi_R$) 
is a  left-moving (right-moving) real fermion field,
and $v$ is the Fermi velocity. 
Depending on the sign of the mass $m$, the gapped phase 
is a topologically trivial/non-trivial phase
(topological/ordinary superconductor)
in symmetry class D in the ``ten-fold way'' classification
of topological insulators and superconductors\cite{Classification2,TenFoldWayNJPhys2010,Kitaev2009}
(To be more precise, 
in order for this topological phase to be stable,  fermion parity needs to be preserved.
We will come back  later to the role of symmetries protecting the SPT phase.)
Which sign of the mass realizes a topological phase 
depends on the ultraviolet (UV) physics 
which is not encoded in the low-energy action.
However, when we make a domain wall in the mass, there is an isolated zero energy Majorana mode
which is insensitive to UV physics.

One can also consider a coupling constant which is
space-dependent, $m \to m(x)$. 
In particular, one can consider a profile where 
$m(x)$ has alternate signs for $x>0$ and $x<0$ as
\begin{align}
m(x)\to \left\{
\begin{array}{ll}
+|m|,  & x\to +\infty
\\
\\
-|m|, & x \to -\infty.
\end{array}
\right.
\end{align}
The mass profile $m(x)$ crosses zero  somewhere 
in between, say at $x=0$.
This geometry realizes an interface 
between
topologically trivial and topologically non-trivial
gapped phases. 
This Jackiw-Rebbi domain wall traps a Majorana fermion,
which is the hallmark of a topological phase. 

One of the purposes of this paper is to extend this Jackiw-Rebbi phenomenon
to interacting settings (see below). In fact,
we will claim that BCFT is a natural language to discuss 
interacting Jackiw-Rebbi phenomena. 
In topological phases, 
details of the profile $m(x)$ do not matter,
and we can make 
the interface between the topological and non-topological phases as smooth as possible. 
If we do so, the transient region
can be made very long 
(compared to the UV cut-off length scale of the theory), and it then looks like a CFT. 
In passing, note that because of topology, the number of stable boundary modes (zero modes)
should not 
change even if 
we make the transient region as long as possible. (See Fig.\ \ref{Fig1}.)

%
The above non-interacting setting can be generalized to more generic, interacting SPT phases and quantum critical points. 
A given gapped phase in (1+1) dimensions can be obtained as a ``massive deformation'' of a CFT,
\begin{align}
S_{*}\to S_{*} -\lambda \int dtdx\, \mathcal{O}(t,x) 
\end{align}
where 
$\mathcal{O}(x)$ is a relevant operator,
and
$\lambda \in \mathbb{R}$ is the coupling constant (see also Ref.\ \onlinecite{ChoLudwigRyuarXiv2016}). 
To consider an interface separating 
trivial and topological phases, 
one can also consider a coupling constant which is
space-dependent, $\lambda\to \lambda(x)$.
(In discussing an interface in free-fermion systems, 
this prescription of creating an interface essentially exhausts all possible interesting cases.
As a working hypothesis, 
we assume this prescription is generic enough even for 
interacting fermion systems.)
In particular, one can consider a profile where 
$\lambda$ has alternate sign for $x>0$ and $x<0$ as
\begin{align}
\lambda(x)\to \left\{
\begin{array}{ll}
+|\lambda|,  & x\to +\infty
\\
\\
-|\lambda|, & x \to -\infty.
\end{array}
\right.
\end{align}
$\lambda(x)$ crosses zero in somewhere 
in between, say at $x=0$.
This geometry realizes an interface 
between
topologically trivial and topologically non-trivial
gapped phases. 
In topological phases, 
details of the profile $\lambda(x)$ do not matter,
and we can make 
the interface between the topological and non-topological phases as smooth as possible. 
If we do so, the transient region
can be made very  long 
(relative to the UV cut-off), andit then  looks like a long region of a  CFT 
described by the action $S_*$. 
In passing, note that because of topology, the number of stable boundary modes (zero modes)
should not
change even if 
we make the transient region as long as possible.

As before, this is precisely the setting of BCFT. 
The CFT realized in the critical region near $x=0$ 
can be viewed as terminated by two (different) gapped phases 
on the left ($x\to -\infty$)
and on the right ($x\to +\infty$).
At low energies, the interface
between the CFT and any of the adjacent gapped phases is expected to 
renormalize
 into a conformal invariant (boundary)  fixed point of the CFT, 
and in this infrared (IR) limit,
the two gapped phases simply look like a two
conformally  invariant boundary conditions of  the CFT. 
In other words, 
this suggests that there is a correspondence 
between gapped topological phases in (1+1) dimensions
and conformal invariant boundary conditions or 
{\it boundary states} in BCFT. 
This point will be further elaborated in the following.

\subsection{Scattering from topological phases}
\label{LabelSubSection-ScatteringFromTopPhases}

\begin{figure}[t] 
\begin{center}
\begin{picture}(240,80)(0,-80) 
\thicklines
\put(  -0,-10){\line(1,0){240}}
\put(  -0,-50){\line(1,0){240}}
%
%
\put( 50,-50){\line(0,1){40}}
\put(190,-50){\line(0,1){40}}
%
%
\thinlines
\put(120,-60){\vector(-1,0){70}}
\put(120,-60){\vector( 1,0){70}}
\put(117,-70){$L$}
\thicklines
\put(50,-20){\line(1,-1){30}}
\put(50,-30){\line(1,-1){20}}
\put(50,-40){\line(1,-1){10}}
\multiput(50,-10)(10,0){11}{\line(1,-1){40}} 
\put(190,-20){\line(-1,1){10}}
\put(190,-30){\line(-1,1){20}}
\put(190,-40){\line(-1,1){30}}
\put(40,-20){\vector(-1,0){25}}
\put(15,-40){\vector(1,0){25}}
\put(225,-40){\vector(-1,0){25}}
\put(200,-20){\vector(1,0){25}}
\put(-5,-20){$\chi^{\mathrm{out}}_{\mathrm{I}}$}
\put(-5,-40){$\chi^{\mathrm{in}}_{\mathrm{I}}$}
\put(230,-40){$\chi^{\mathrm{out}}_{\mathrm{II}}$}
\put(230,-20){$\chi^{\mathrm{in}}_{\mathrm{II}}$}
\end{picture}
\end{center}
\caption{
\label{fig:S-matrix}
Scattering from a (1+1)-dimensional SPT phase (shaded region).
$\chi^{\mathrm{in}/\mathrm{out}}_{\mathrm{I}/\mathrm{II}}$
represent
the amplitudes of the in-coming/out-going single-particle states 
in Region I/II (each of these amplitudes  being an $N$-dimensional vector representing $N$ channels). 
        }
\end{figure}
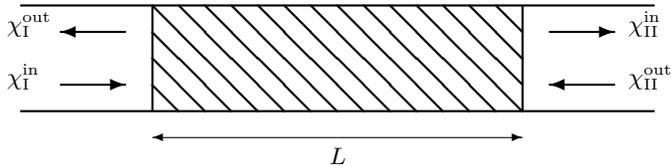

The second argument relating SPTs and BCFTs 
is motivated by 
the scattering matrix formulae of topological invariants of free fermion SPTs. 
In Ref.\ \onlinecite{Akhmerov2011},
properties of the scattering matrix that 
describes scattering of free fermion states (in the ``ideal lead'')
off a given topological phase were discussed. 
As an example, let us consider a quasi-1d system 
and use  the following construction: 
we connect two $2 N$-channel wires   (in the Majorana basis) 
to the two sides of a quasi-1d scattering region,
which is a gapped phase (see FIG.\ \ref{fig:S-matrix}).
We are after the topological properties of the gapped region. 
This situation can be modeled by the following single-particle Hamiltonian
\begin{align}
\mathcal{H}
=
- {i} \frac{d}{dx}\sigma_3\otimes I_N
+V(x),
\end{align}
where $V(x)$ is a potential. 
For our purpose,
$V(x)= m \sigma_2 \otimes I_N$
inside the gapped region, whereas
$V(x)=0$ in the lead. 
The single-particle Hamiltonian satisfies the particle-hole constraint,
$\mathcal{H}^* = -\mathcal{H}$,
and belongs to symmetry class D. 
Let us consider 
an asymptotic state with energy $\varepsilon=k$ entering
the scattering region $[0,L]$ of  length
$L$ located  to the right of  $x=0$
with  amplitudes $\chi^{\mathrm{in}}_{\mathrm{II}/\mathrm{I}}$
\begin{align}
\Psi^{\mathrm{in}}(x)=
\left\{
\begin{array}{lc}
\displaystyle \chi^{\mathrm{in}}_{\mathrm{II}}e^{-{i}k(x-L)}n_{-}, & L< x, \\
& \\
\displaystyle \chi^{\mathrm{in}}_{\mathrm{I}}e^{+{i}kx}n_{+}, & x <0,
\end{array}
\right.
\label{eq: basis for asymptotic in state}
\end{align}
and a scattered state
emerging from the disorder potentials 
with the same energy $\varepsilon$ and 
the amplitudes $\chi^{\mathrm{out}}_{\mathrm{II}/\mathrm{I}}$
\begin{align}
\Psi^{\mathrm{out}}(x)=
\left\{
\begin{array}{lc}
\displaystyle \chi^{\mathrm{out}}_{\mathrm{II}}e^{+{i}k(x-L)}n_{+}, & L< x, \\
& \\
\displaystyle \chi^{\mathrm{out}}_{\mathrm{I}}e^{-{i}kx}n_{-}, & x <0,
\end{array}
\right.
\label{eq: basis for asymptotic out state}
\end{align}
where the (column) vectors
$n_{\pm}$ are given\footnote{The superscript ${}^T$ denotes the transpose}  by 
$n^{\ }_{+}:=(0, ..., 1, ...;0, ...,0)^{{T}}$,
$n^{\ }_{-}:=(0,..., 0; 0, ...,1,...,0)^{{T}}$.
Note that $e^{+{i}kx}n_{+}$ 
and $e^{-{i}kx}n_{-}$
are for $k>0$,
a right-moving and a left-moving wave function, respectively,  since
the  eigenvalue of the momentum operator $-{i}{d}/{d}x$ 
is positive (negative). 

The $2N \times 2N$  scattering matrix relates incoming and outgoing amplitudes in the two regions I and II as
\begin{align}
\label{LabelEq-DEF-ScatteringMatrix}
&
\left(
\begin{array}{c}
\chi^{\mathrm{out}}_{\mathrm{I}} \\
\chi^{\mathrm{out}}_{\mathrm{II}} \\
\end{array}
\right)
=
S
\left(
\begin{array}{c}
\chi^{\mathrm{in}}_{\mathrm{I}} \\
\chi^{\mathrm{in}}_{\mathrm{II}} \\
\end{array}
\right),
\quad
S(\varepsilon) =
\left(
\begin{array}{cc}
r(\varepsilon)& t'(\varepsilon)
\\
t(\varepsilon) & r'(\varepsilon)
\end{array} 
\right),
\end{align}
where $r (r')$ and $t (t')$ are $N \times N$ matrices representing the reflection part and transmission part of the scattering matrix. 
Here we use the (standard)  convention 
 that $r$ and $t$ 
describe the
reflection and transmission coefficients of the incoming states from left hand side ($\chi_I^{\rm in}$), 
while $r'$ and $t'$ 
describe the reflection and transmission coefficients  of the incoming states from right hand side ($\chi_I^{\rm in}$)
- compare FIG. \ref{fig:S-matrix}.

Topological properties of the scatterer are fully encoded in, and can be read off from the $S$-matrix as follows. 
Since the scatterer (i.e., a gapped (1+1)d phase) is gapped, if $L$ is large enough,
(almost)
all\footnote{All, when $L=\infty$}  incoming electrons eventually get reflected back from the scatterer. 
We can thus focus on the reflection part, $r$, of the $S$-matrix. 
Depending on the underlying symmetry of the problem,  
the reflection matrix is subject to a set of constraints. 
For symmetry class D, for example, 
the space of reflection matrices (at $\varepsilon=0$),
denoted by $\mathcal{R}$,
is disconnected, $\pi_0(\mathcal{R})=\mathbb{Z}_2$,  
which corresponds to the $\mathbb{Z}_2$ classification of
class D in (1+1) dimensions. 
These two sectors are distinguished by the $\mathbb{Z}_2$-valued
topological index,
\begin{align}
\mathrm{sgn}\, \mathrm{det}\, r (\varepsilon=0)
=
\pm 1.
\end{align}
Here, 
when 
$\mathrm{sgn}\, \mathrm{det}\, r 
=
1
$
the gapped system attached to the lead is trivial.
On the other hand, 
when
$
\mathrm{sgn}\, \mathrm{det}\, r 
=
-1
$
the gapped system attached to the lead is non-trivial.
In this way, the topological character of the bulk is fully 
encoded in the scattering matrix. 
\footnote{
Similarly, the transmission can detect
the quantum phase transition separating 
trivial topological phases.}

If we further impose, for example,  time-reversal symmetry which squares $+1$,  
the relevant symmetry class is class BDI of the  Altland-Zirnbauer classification. 
The topological classification at the level of non-interacting fermions is given in terms of a integral valued topological invariant
(``the winding number''). 
In the scattering matrix approach,
the integer topological invariant is given by the number of negative eigenvalues of 
the reflection matrix $r$. 
(In passing, we note that this non-interacting topological invariant 
fails to capture 
the
reduction  of the non-interacting classification from $\mathbb{Z}$ to $\mathbb{Z}_8$
found by Fidkowski and Kitaev
\cite{Fidkowski2010, fidkowski2010entanglement} in the presence of interactions.)

The construction described above is precisely the typical setting of BCFTs.
The gapless ideal lead that we use to detect topological properties of the SPT phase
is a special case of a  CFT, which has its boundary  condition set by the SPT. 
In BCFTs the boundary can be probed by correlations of bulk fields.
For example, 
BCFT computes the (right-left) fermion two-point function in the presence of a boundary, 
which is given by
\cite{LudwigAffleck1991,
AffleckLudwig1993,
CardyLewellen1991,
Maldacena1997}
\begin{align}
\langle \psi_L^a(z) \psi^b_R(\bar{z})\rangle =
\frac{r_{ab}}{z-\bar{z}}. 
\end{align} 
Here, the physical spacetime
consists of 
the upper half  complex plane,
and the boundary is located 
on the real axis,  $z=\bar{z}$.
In the absence of interactions, the amplitude $r$ of this function  contains the information about 
the single-particle scattering matrix
(the only existing scattering matrix in the absence of interactions). 
When the topological winding number is zero,
the single particle Green function in the CFT is given by
$\mathrm{sgn}\, \det r=1$.
On the other hand,
when the topological winding number is non-zero, 
$\mathrm{sgn}\, \det r=-1$.

The above consideration shows that  
a CFT can be used as an external ``probe'' to look into possible topological bulk states,
although the framework presented so far has been limited to non-interacting fermion systems. 
However, BCFTs in general are not limited to non-interacting systems 
and are expected to give us a framework to study interacting (1+1)d SPT phases in general.
The reason why our consideration so far is limited to free-fermion systems 
is the fact that we focused on the single-particle $S$-matrix, or 
the single-particle fermion Green's function $\langle \psi^a_L(z) \psi^b_R(\bar{z})\rangle$. 
For non-interacting problems,
unitarity restricts $|r|=1$. 
In the presence of interactions,
however,  even if they only act on the boundary,
it is not difficult to find examples
of boundary conditions
where $|r|<1$, and in particular, we have examples where
$r=0$. 
This is known for example in the context of the two-channel Kondo and related models.\cite{Maldacena1997}
In these interacting systems, unitary of the $S$-matrix can be violated within the single-particle
sector (while   unitary in the full many-particle
Hilbert space is of course preserved).
BCFTs are not limited to the description of 
the single-particle fermion Green's function $\langle \psi(z) \psi(\bar{z})\rangle$,
but give us the description of the full (`many-body', or `Fock-')  Hilbert space in the presence of interactions, 
even if
 the interactions are only  operative within the gapped region
(i.e., only at the boundary of the ideal lead (CFT)). 
In Sec.\ \ref{BDI topological superconductors in (1+1)d},
we will show that our approach based on  BCFTs indeed yields the $\mathbb{Z}_8$ classification of Fidkowski-Kitaev
in the presence of interactions. In that section, the CFT (``in the lead'') is taken to consist of non-interacting
Fermions, while all interactions occur solely on the boundary. In the language of the {\it entanglement spectrum}, to be 
discussed  in the following subsection \ref{LabelSectionEntanglementSpectrum}, this corresponds physically to a situation of a quantum phase transition  out of the interacting SPT phase
into a trivial phase, described by non-interacting massless Fermions. Since, as will be described in the next subsection, one of the boundaries of the BCFT describing the entanglement
spectrum corresponds to an interface of the CFT (in the present case a non-interacting theory) with the fully interacting SPT phase,
all interactions are incorporated into that boundary condition.

\subsection{The entanglement spectrum}
\label{LabelSectionEntanglementSpectrum}

The last argument in this section is based on the entanglement Hamiltonian and the entanglement spectrum of SPT phases.
This is the most general and most fundamental of the arguments we are giving. As we will see, it will ``automatically'' choose for us a
gapless CFT, and a suitable BCFT. As we will now explain, the entanglement Hamiltonian of the SPT plays the role
of the ``expanded domain wall'' or the ``lead'' of the previous two subsections.

Vital tools for the  study of
one-dimensional (and other) gapped phases
are the entanglement entropy and
the entanglement spectrum.
In gapped (1+1) d phases which are adjacent to a CFT with central charge $c$,
i.e. in which the correlation length $\xi$ is much larger than the  microscopic length $a$ ('scaling limit'),
it is well known that the entanglement entropy behaves as 
\begin{equation}\label{EntanglementEntropyConstantTopology}
S_A 
\simeq 
({c}/{6})\ln(\xi /a)
+
\mbox{constant.} 
\end{equation}
A topological phase, being gapped, can be tuned to have 
a minimally 
("infinitely")
 short correlation length, $\xi=a$. For such a representative of the topological phase
only the constant term  in \eqref{EntanglementEntropyConstantTopology} remains. 
Because it is a topological phase, it is not possible to make
the entanglement entropy vanish completely,
while preserving the symmetry which protects the  SPT phase under consideration.
This non-vanishing constant part (the part which is not controlled by the correlation length) 
is a key to classifying gapped phases in one spatial dimension.
(In fact, this is the part that a matrix product state is capable of capturing).  

\begin{figure}
\begin{center}
\includegraphics[width=\columnwidth]{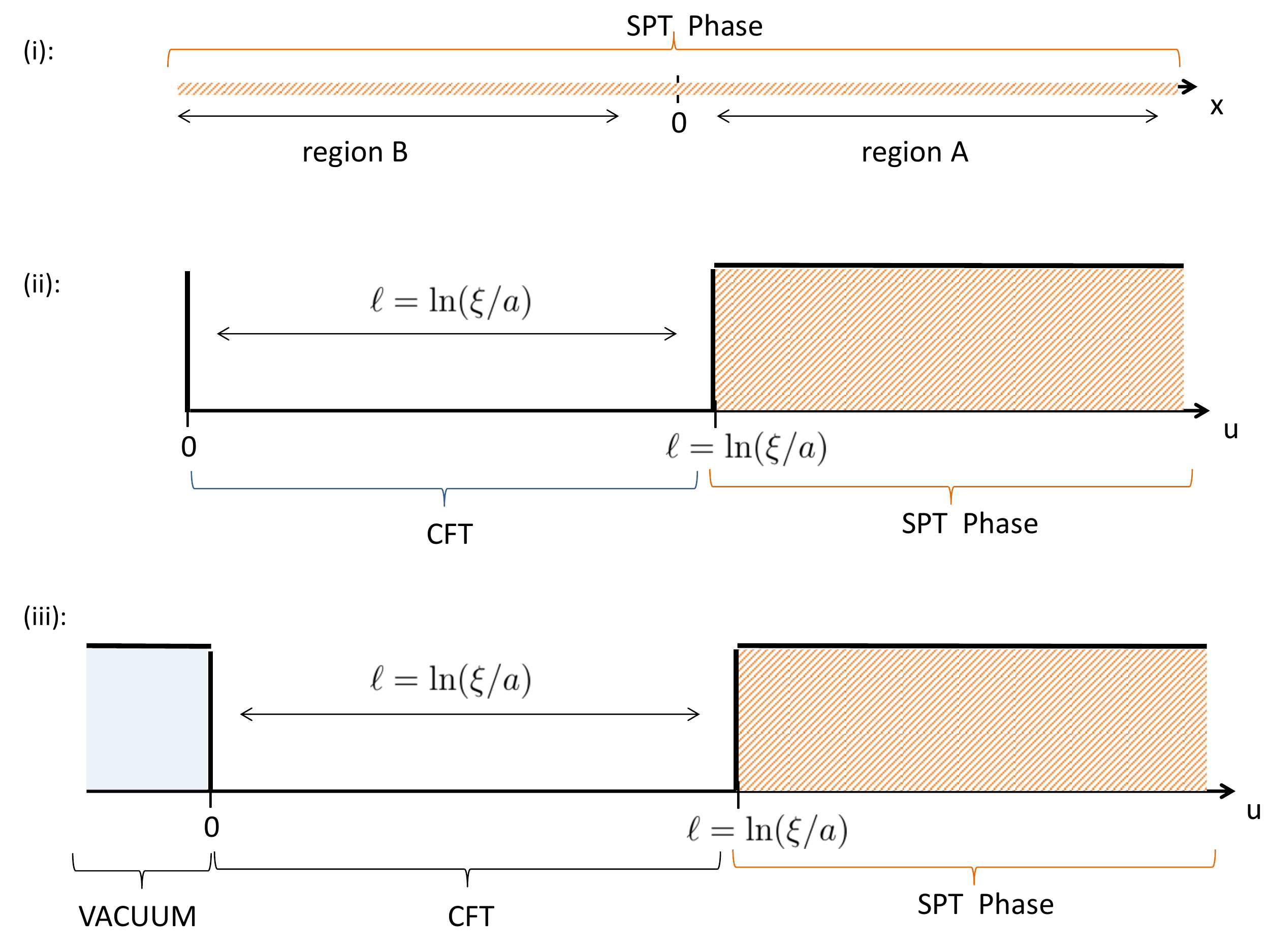}
\caption{Entanglement Hamiltonian of SPT phase (from Cho et al., arXiv:1603.04016). (i): The ground state of the (1+1) dimensional SPT phase of correlation length $\xi$,
 in the vicinty of a quantum
critical point  on the infinite space $-\infty < x < +\infty$, bipartitioned
into region A (positive $x$) and region B (negative $x$).  (ii): The entanglement Hamiltonian (defined on a space with coordinate $u$ which
is different from $x$)  is that of  the CFT describing the quantum critical
point, but confined to a finite interval of length $\ell= \ln (\xi/a)$. On the right hand side of the interval is an interface of the CFT with the
gapped SPT, providing one boundary condition. On the other side of the interval the CFT simply ends, providing another
boundary conditions  (``free boundary condition'').
(iii): The situation depicted in (ii) is a generalization of the ``expanded Jackiw-Rebbi domain wall''  depicted in FIG. \ref{Fig1}  (c),
 to the case of a completely general interacting SPT.
}
 \label{LabelFigEntanglementHamiltonian}
\end{center}
\end{figure}

Much more information about the SPT phase is contained in the entanglement spectrum.
Indeed, it was recently shown in Ref.\ \onlinecite{ChoLudwigRyuarXiv2016}
that the entire  low-lying entanglement spectrum of a gapped  phase close
a  quantum critical point,  such as the SPT under consideration,  is  universal and described, very generally, by 
the CFT describing the quantum critical point itself, but  on  a finite interval of length $\ell=\ln(\xi/a)$ with
 suitable boundary conditions. In short, the entanglement spectrum is described by a
boundary conformal field theory. In particular, the boundary
condition  at one end of the finite interval is  determined by the specific gapped phase in the vicinity of the quantum critical point;
different boundary conditions correspond in general to  different gapped phases adjacent to the same quantum critical point.
In other words, there is a  mapping between gapped phases in the vicinity of the quantum
critical point, and boundary conditions on the CFT on a finite interval which describes  the entanglement Hamiltonian.
This property  is the key to relate the gapped SPT to a (gapless) boundary CFT. 
In fact, this  materializes
in complete generality the connection, discussed above, between the gapped (1+1) dimensional  SPT
phase and the gapless boundary CFT (see FIG. \ref{LabelFigEntanglementHamiltonian}).

From this point of view, the constant part of the entanglement entropy in \eqref{EntanglementEntropyConstantTopology}
should come from the fact that we have
to specify particular boundary conditions
on the CFT, in order for the resulting boundary CFT to represent the
entanglement
Hamiltonian of the SPT.
I.e.
the constant part of the entanglement entropy is related to the boundary states
in a given CFT.

Roughly speaking, the classification problem of 1d gapped phase is thus related
to the classification of the boundary states in CFT.
Since we take the  $\xi\to a$ limit, 
the constant part of the entropy is given by an overlap
between two boundary states. 
This is somewhat reminiscent of the Affleck-Ludwig boundary entropy\cite{Affleck:1991tk},
except that here  we have to 
consider an opposite  limit (for details see below).

Specifically, making a  connection with the  setting  discussed in the previous section,
i.e., with  the scattering off of SPT phases, 
the entanglement Hamiltonian (somewhat surprisingly) 
realizes precisely the same setting, but in a completely general context. 
Note that while
 in the previous setting, there may be an ambiguity 
as to our choice of ideal leads, 
 in the entanglement Hamiltonian,
on the other hand,
the lead (i.e., the CFT) is ``automatically'' chosen. 

As a side remark, 
a  connection between gapped topological phases and critical systems can be also made 
by following the construction in Ref.\ \onlinecite{HsiehFuQi2014} of the 
so-called bulk entanglement spectrum.
In Ref.\ \onlinecite{HsiehFuQi2014}, 
it was shown that  for an SPT, 
by using a biparition of position space into regions A and B with the property that the interface
between A and B grows with the volume (in 1D with length),   a gapless
entanglement Hamiltonian can emerge. Concretely, 
using a MPS construction of a gapped SPT phase (e.g., the Haldane phase), 
and  rearranging tensors in the MPS in a 
staggered way gives rise to a transfer matrix of a critical system (the six-vertex model),
describing a spin-1/2 Heisenberg spin chain. We can think of this construction as generating
an entanglement Hamiltonian which sits at
the quantum phase transition out of the Haldane phase into the dimerized phase
of the spin-1  chain driven by staggering. This quantum phase transition is in the universality class of
the unstaggered spin-1/2 Heisenberg chain, describing the CFT of the (gapless) entanglement Hamiltonian.

\section{Symmetry-protected degeneracy in entanglement spectrum: general considerations}
\label{LabelSection-SymmetryProtectedDegeneracyEntanglementSpectrumGeneralConsiderations}

Matrix product states (MPSs)
provide  a convenient framework to discuss 
entanglement 
and  in particular the entanglement spectrum of  gapped phases in (1+1) dimensions. 
In particular, 
the symmetry protected degeneracy of the entanglement spectra of  SPT phases
-- a hallmark of SPT phases --
can be understood from the MPS perspective.  
\cite{Ryu2006, Pollmann2010}
In this section, we will  understand  
the symmetry protected degeneracy of the entanglement spectra 
of SPT phases
from the perspective of continuous field theories,
in particular orbifold BCFTs, 
in order to provide an alternative point of view.

Specifically, in this section we consider the BCFT describing the entanglement Hamiltonian, as discussed in the preceding section
\ref{LabelSectionEntanglementSpectrum}: the entanglement spectrum is that of the CFT describing the quantum phase transition
itself,
but on an interval of length $\ell=\ln(\xi/a)$ with two boundary conditions A and B at the two ends. In this section we focus
on the requirements on these boundary conditions, which  arise from the fact that we are describing the entanglement
spectrum of a SPT which is protected by a symmetry group $G$.

\subsection{Quick review of  Boundary Conformal Field Theory (BCFT)}

From the discussion in the previous section,
we associate a particular BCFT 
(or a particular {\it boundary condition}, and {\it boundary state} of a CFT) with a given SPT phase;
different SPT phases in proximity of the same  quantum critical point 
are described by 
different boundary conditions on  the same  CFT. 
(More precisely, the correspondence between a BCFT and a SPT phase is not one-to-one,
but rather 
several
different BCFTs may correspond to a given SPT phase, because there may be several different
quantum critical points through which one can exit a given SPT phase into other, neighboring phases. By staying close to a 
particular quantum  critical point, we pick a particular BCFT-description of a given SPT phase.)

Let us first briefly  review the general framework of BCFTs.
Consider a CFT on a finite spatial interval with boundary conditions specified by $A$ and $B$
at the two boundaries.
In BCFT, we may  compute the partition function by 
using one of the following two alternative pictures:
\cite{Cardy1989}
the so-called {\it open string}
picture (sometimes also called {\it loop channel} picture), and the so-called
{\it closed string}
picture (sometimes also called {\it tree channel} picture).
First, in the open string
picture, the partition function (at inverse temperature $\beta$)  is  written  as a trace
of the Hamiltonian ${\hat H}_{AB}^{open}$ of the finite interval of length $\ell$ with
boundary conditions $A$ and $B$ at the two end points (boundaries) of the interval:
\begin{align}\label{OpenStringPartitionFunctionZAB}
Z_{AB} 
&= \mathrm{Tr}_{\mathcal{H}_{AB}}\, e^{- \beta {\hat H}^{open}_{AB} } 
\nonumber \\
&=
\mathrm{Tr}_{\mathcal{H}_{AB}}\, 
q^{ \hat{H}_L}.
\end{align}
Here $\mathcal{H}_{AB}$  denotes the Hilbert space of quantum states on the interval.
In the second line, 
the partition function is rewritten, by using the 
"folding procedure", so that say
only holomorphic (left-moving) degrees of freedom appear\footnote{Owing to its assumed conformal invariance, at  each boundary right movers can be viewed
as left movers,  analytically continued into the lower half complex plane};
here ${\hat H}_L$ is the Hamiltonian
defined purely in the holomorphic sector
where it  can be expressed   in terms of the (holomorphic)  Virasoro generator ${\hat L}_0$ and the central charge $c$
as ${\hat H}_L = {\hat L}_0 - c/24$.  
All terms in the partition function in Eq.\ (\ref{OpenStringPartitionFunctionZAB}) are powers of
\begin{align}
q = e^{ - \pi \beta/\ell}
\end{align}
related to the length ($=\ell$)  of the system  ($2\ell$ for the left-movers after `folding')  and the inverse temperature $\beta$.

The structure of the "open string"  Hilbert space $\mathcal{H}_{AB}$
depends on the choice of the boundary conditions $A$ and $B$.
In particular, the Hilbert space 
$\mathcal{H}_{AB}$ can be decomposed into 
different irreducible representations $\phi_a$ of the Virasoro (or more generally,  a larger "chiral") algebra of the CFT,
which is supported on a vector space  $[\phi_a]$,
\begin{align}
\mathcal{H}_{AB}= \bigoplus_a \  n^a_{AB} \  [\phi_a] . 
\end{align}
The non-negative integers $n^a_{AB}$ represent the multiplicity with which the irreducible representation  $\phi_a$ occurs.
Hence,  the partition function can be written in the form 
\begin{align}\label{PartitionFunctionOpenStringChannel}
Z_{AB} = \sum_a \ n^{a}_{AB} \ \chi_a(q),
\end{align}
where $\chi_a(q)$ is the partition function associated with  the
representation $\phi_a$ (and is usually called its ``character''). 

The partition function can also  be computed, alternatively, by exchanging the roles of 
the space and imaginary (Euclidean)  time coordinates. 
In the resulting closed  string 
picture,  the partition function can be written in terms of boundary states 
$|A\rangle$ and $|B\rangle$ as
\begin{align}\label{ZABPartitionFunctionClosedStringPicture}
Z_{AB} &= \langle A| e^{ - \ell {\hat H}^{closed} } |B\rangle 
\nonumber \\
&=
\langle A| \tilde{q}^{ \frac{1}{2} ({\hat H}_L+{\hat H}_R) } |B\rangle,
\end{align}
where
${\hat H}^{closed}$ is the Hamiltonian of the CFT on a space with periodic boundary conditions (a circle)
of circumference $\beta$, acting on the corresponding Hilbert space ${\cal H}^{closed}$, which 
is contained in the tensor product of 
holomorphic (left-moving) and anti-holomorphic (right-moving) degrees of freedom. 
The boundary states can be expanded in terms of so-called  {\it Ishibashi states} as
\begin{align}
|A\rangle = \sum_a A_a |a\rangle\!\rangle,
\quad
\langle A| = \sum_a \tilde{A}_a \langle\!\langle a|.
\end{align}
The  Ishibashi states are special states in the closed string Hilbert space ${\cal H}^{closed}$ in which the  holomorphic (left-moving)
and the anti-holomorphic (right-moving) basis states of the Hilbert space are maximally entangled
\footnote{See Ref.\ \onlinecite{qi2012general} for
a discussion from the entanglement perspective.}.
This leads to the second line of Eq.\ (\ref{ZABPartitionFunctionClosedStringPicture}) above,
where
$
{\hat H}_L+{\hat H}_R={\hat L}_0+\hat{\bar{ L}}_0 -c/12
$
is the Hamiltonian of the CFT
with periodic boundary conditions (on the unit circle)
and
\begin{align}
\tilde{q} = e^{-\frac{4\pi \ell}{\beta}}, 
\end{align}
 implying
\begin{equation}
\langle\!\langle a| 
\tilde{q}^{\frac{1}{2} (\hat{L}_0 +\hat{\tilde{L}}_0 -c/12) } | b\rangle\!\rangle
= \ 
\delta_{ab} \ 
\chi_a(
\tilde{q}=
(
e^{-\frac{2\pi \ell}{\beta}})^2
).
\end{equation}
Then, the partition function can be written as
\begin{align}\label{ClosedStringPartitionFunctionZAB}
&
Z_{AB}
=
\sum_a \tilde{A}_a B_a  \chi_a (\tilde{q}). 
\end{align}

The two representations, Eq.\ (\ref{OpenStringPartitionFunctionZAB})
and Eq.\ (\ref{ClosedStringPartitionFunctionZAB})
of the same partition function 
are related by a  modular transformation of the space-(imaginary)time torus:
By using the modular $S$-matrix,  
\begin{align}
\chi_a(q) = \sum_{b} S_a^b \chi_b(\tilde{q}), 
\end{align}
one sees that
the integer coefficients $n^a_{AB}$
in Eq.\ (\ref{PartitionFunctionOpenStringChannel}),
and the expansion coefficients $A_a, B_a$
in Eq.\ (\ref{ClosedStringPartitionFunctionZAB}),
 are related via 
\begin{align}
\label{CardyFormulaModular}
&
\sum_a n^a_{AB} S^{b}_a = \tilde{A}_b B_b,  
\quad 
\mbox{and}
\quad 
n^b_{AB}= \sum_{a}\tilde{A}_a B_a S^b_a.
\end{align}
In the limit $\tilde{q}\to 0$ ($q\to 1$), 
we have the Affleck-Ludwig boundary entropy:
\cite{Affleck:1991tk}
\begin{align}
\nonumber
\log Z_{AB} &=\log \left(
\sum_a n^a_{AB} \ {S_a}^b \ \chi_b(\tilde q) \right)\\
&\sim 
\log \left(\sum_a n_{AB}^a S^0_a\right)
+
\log
\chi_0(\tilde{q})
\nonumber \\
&
\sim \log \tilde{A}_0 +\log B_0 + \log 
{\chi}_0(\tilde{q})
\nonumber \\
&
\to   \log \tilde{A}_0 +\log B_0. 
\end{align} 
(In the last line, use was made of  $\lim_{{\tilde q}\to 0} {\chi}_0 (\tilde{q}) =1$.)

\subsection{Quick review of orbifold CFTs}

A natural and general  framework to discuss the  action of discrete symmetries in CFTs is the so-called orbifold CFT.
\cite{Dijkgraaf1989, Sharpe2003}
In order to discuss BCFT in the context SPT phases, we need to discuss the notion of the  orbifold in BCFT.
\cite{Sharpe2003}
First, before discussing the orbifold of a CFT with boundaries 
(i.e. of BCFT), we give a
very brief overview of orbifold CFTs in the bulk (i.e. on a space with periodic boundary conditions -  in the absence of boundaries). 
\cite{Douglas1998, Douglas1999}
Orbifold CFTs can be obtained from a parent CFT by modding out (``gauging'') a discrete symmetry group $G$.
The partition function of an orbifold CFT on a torus is known to have  the following structure,
\begin{align}
\label{LabelEqBulkOrbifoldPartitionFct}
Z=\frac{1}{|G|} \sum_{g,h\in G}^{[g,h]=e}\,
\varepsilon(g|h) \ 
Z(g,h),
\end{align}
where $Z(g,h)$ denotes the partition function in the sector  twisted by  group elements
$g$ and $h$ in the (imaginary)  time and space directions, respectively (see below).
Here  $[g,h]=ghg^{-1}h^{-1}$ denotes the commutator in the group.
The sector-dependent phases, $\varepsilon(g|h)$, are called discrete torsion, and will be defined in detail below. 
\cite{Vafa1986}
In each sector,
the (bulk) partition function 
is given by
\begin{align}
Z(g,h) =\mathrm{Tr}_{\mathcal{H}_h}
\left [
\hat{g} \  q^{{\hat H}_L} \bar{q}^{{\hat H}_R}
\right ] 
=
\sum_{(j,\bar{j})} \chi^g_{h,(j)} (q) \bar{\chi}^g_{h,(\bar{j})} (\bar{q}). 
\end{align}
Here, $\mathcal{H}_h$ is the Hilbert space of the sector 
twisted\footnote{Both right- and left-moving factors of the bulk Hilbert space are twisted by the same group element $h$.} 
by $h$.
Each twisted-sector Hilbert space $\mathcal{H}_h$ 
is decomposed into irreducible  representations  (denoted by $(j)$ and $(\bar{j}$))
of the  
left- and right-moving Virasoro (or possibly of some larger chiral\cite{Dijkgraaf1989})
algebra, and 
we introduced the corresponding chiral blocks (``characters'')
\begin{align}
\chi^g_{h,(j)}(q) = \mathrm{Tr}_{\mathcal{H}_{h,(j)}} \left[
\hat{g} \  q^{{\hat H}_L}
\right]. 
\end{align}
Here $\hat{g}$ is a representation of the group element $g \in G$
on the Hilbert space $\mathcal{H}_{h, (j)}$. (Note that for each group element $h$, the sum over group elements
$g$ commuting with $h$ (i.e.: $[g,h]=e$)
in the total partition function,
Eq.\ (\ref{LabelEqBulkOrbifoldPartitionFct}),
projects
onto $N_h$-invariant states where
\begin{align}
N_h = \{ g\in G| [g,h]=e\}
\end{align}
is the `normalizer' of $h$.)

\subsection{Symmetry-protected degeneracy} 

After the above review of general BCFT, and of bulk orbifold CFTs, 
we will now discuss the symmetry-protected degeneracy of 
the entanglement spectrum of an SPT phase.
Let us start by observing 
that the multiplicity coefficients $n^a_{AB}$ appearing in
Eq.\ (\ref{PartitionFunctionOpenStringChannel}) for $Z_{AB}$
are closely related to the symmetry-protected degeneracy:
All states in the representation $a$ of the Virasoro (or larger chiral) algebra are at least $n^a_{AB}$-fold degenerate.
In particular, the ground state in each representation $a$ appearing in $Z_{AB}$  is $n^a_{AB}$-fold degenerate,
which can be seen by taking the limit $q\to 0$ ($\tilde{q}\to 1$).
In this limit, the partition function behaves as 
$
Z_{AB} \sim \sum_a n^a_{AB} q^{-c/24 + h_a}
$
where $h_a$ is the lowest energy state in a given Virasoro representation $a$,
which we assume non-degenerate. 
Observe that the multiplicity $n^a_{AB}$ 
can be extracted by taking the limit $q\to 0$ , which is opposite to the limit $\tilde{q}\to 1$ taken in the Affleck-Ludwig boundary entropy. 
As in the boundary entropy, 
one can express the degeneracy $n_{AB}^a$ in terms of the data constituting the boundary states
as follows by taking the limit  $q\to 0$
 (using the second equation in Eq.\ (\ref{CardyFormulaModular})) :
\begin{align}\label{ZABDegeneracy}
\log Z_{AB} &= 
\log \left (\sum_a n_{AB}^a \chi_a(q) \right )
\nonumber \\
&\sim \log \left ( n_{AB}^0 \chi_0(q) \right ) \nonumber \\
&\sim \log (n^0_{AB})
\end{align}
Thus,  as far as 
the identity representation ``$0$"
appears in $Z_{AB}$,
$n^0_{AB}= \sum_a \tilde{A}_a B_a S^0_a$
yields  the degeneracy.

While the 
multiplicities  $n^a_{AB}$ 
are to be closely related to 
the symmetry-protected degeneracy, 
in the above discussion we have not mentioned symmetry at all.   
In the following, 
we will be interested in the situation where the multiplicity $n^a_{AB}$ results from a
discrete symmetry of the BCFT, and if so, we are interested in relating it to the property of 
the boundary conditions set by SPT phase. 
In other words, the multiplicity (degeneracy) could be simply an accidental one.
On the other hand, if the SPT phase of  interest is topologically non-trivial,
we expect $n^a_{AB}>1$ is enforced by symmetry. 
We want to be able to understand the multiplicity (degeneracy) as arising from the symmetry that
protects the SPT Phase.

\subsection{Projective representation in the open string channel} 

Coming back to the open string picture,
Eq.\ (\ref{OpenStringPartitionFunctionZAB}),
the partition function can be written as a chiral block (as discussed above):
\begin{align}\label{OpenStringPartitionFunctionUnprojected}
 Z_{AB} = 
\mathrm{Tr}_{\mathcal{H}_{AB}}\,
\left [
 e^{- \beta {\hat H}^{open}_{AB} }
\right ]
 =
\mathrm{Tr}_{\mathcal{H}_{AB} }\, q^{\hat{H}_L}
\end{align}
(i.e., we used the
'folding procedure'  to write  the partition function 
purely in terms of the chiral (left-moving) sector of the theory.)
The trace here is taken with respect to the Hilbert space $\mathcal{H}_{AB}$,
which is determine by boundary conditions $A$ and $B$.  
As in a typical set-up of orbifold CFTs, 
\cite{Dijkgraaf1989} 
we assume a decomposition of the Hilbert space 
of the form 
\begin{align}
 \mathcal{H}_{AB} = \bigoplus_{a} \ 
r_{a}
\otimes
[\phi_{a}],
 \label{decompose}
\end{align}
where 
 $r_{a}$
and
$[\phi_{a}]$ 
denote an irreducible representation of the
 finite group  $G$
and of the
Virasoro (chiral)  algebra, respectively. 
Then, the partition function can be written as
\begin{align}\label{UntwistedOpenStringPartitionFunction}
Z_{AB}=\sum_{a} \rho_{a}(1) \chi_{a}(q),
\end{align}
where $\rho_{a}(g)$ is the group character of the irreducible representation $r_{a}$
evaluated on the group element  $g\in G$.
In this description, the degeneracy factor 
from Eq.\ (\ref{PartitionFunctionOpenStringChannel}) appears in the form
\begin{equation}\label{RhoOneEqualsMultiplicity}
\rho_{a}(1)=\mathrm{dim}\, r_{a}=n^a_{AB},
\end{equation} 
and is attributed to the invariance of the Hamiltonian 
under the symmetry group $G$ and to the appearance of representations of $G$ 
of dimension larger than one in the spectrum.

We now consider a slight generalization of the partition function $Z_{AB}$ 
written in  Eq.\ (\ref{OpenStringPartitionFunctionUnprojected}) above, 
namely the `open-string orbifold partition function' defined by
\begin{align}
\label{LabelEqOpenStringOrbifoldPartitionFunction}
Z^{orb}_{AB}
=
|G|^{-1}
\sum_{g\in G}
\mathrm{Tr}_{\mathcal{H}_{AB} }\, 
\left[
\hat{g} e^{-\beta \hat{H}^{open}_{AB}}
\right].
\end{align}
Here, we denote by $\hat{g}$
the representation of the group element $g$ on the Hilbert space
$\mathcal{H}_{AB}$.
With the decomposition \eqref{decompose}, 
$\hat{g}$ can be decomposed accordingly into irreducible components as
\begin{align}\label{LabelEqghat}
\hat{g}=\bigoplus_a D_a (g) 
\end{align}
where $D_a(g)$ is the representation matrix of $g$ in the irreducible representation $r_a$.
If we think of the finite interval of length $\ell$ on which the BCFT resides from the point of view of the expanded domain
wall picture of Sec.\ \ref{LabelSectionExpandedDomainWallPicture}, we see that for small size $\ell$ the gapless
BCFT region reduces to the local domain wall at which we expect to see the appearance of a projective representation of the
symmetry group $G$ defining the SPT phase. Therefore, we expect to see a projective representation of the symmetry group
on the Hilbert space ${\cal H}_{AB}$, since this just describes the expanded version of the domain wall 
(Sec.\ \ref{LabelSectionExpandedDomainWallPicture}).
Therefore we
will be interested in the possible appearance of {\it projective} representations of the group $G$, 
for which
the representation matrices 
$D_a$ will in general satisfy the composition law
\begin{align}
D_a(g)D_a(h) = \omega(g|h) D_a(gh),
\label{proj. rep}
\end{align}
where  $g,h\in G$, and 
where $\omega(g|h)$ is a two-cocycle in the  cohomology group  $H^2(G, U(1))$.
Note that in the direct sum decomposition in \eqref{decompose}, 
all representations $r_a$ should have the same two-cocycle.
(In general, one can take a direct product of two representations 
having different two-cocyles, but not a direct sum thereof. )

Using the decomposition \eqref{decompose}, the orbifold partition function 
in Eq. (\ref{LabelEqOpenStringOrbifoldPartitionFunction})
can be expressed in terms of the ``twisted partition functions''
\begin{align}\label{TwistedOpenStringPartitionFunction}
Z^g_{AB}=
\mathrm{Tr}_{\mathcal{H}_{AB} }\, 
\left[
\hat{g} e^{-\beta \hat{H}^{open}_{AB}}
\right]
=
\sum_{a} \rho_{a}(g) \chi_{a}(q),
\end{align}
where $\rho_a(g)=\mathrm{tr}\, D_a(g)$ defines the character of a representation $D_a$ in the usual manner.
The twisted partition function $Z^g_{AB}$ thus extracts the characters of the representations of $G$.  
This  twisted partition function may then be used to identify the representation
${\hat g}$  
appearing in the {\it untwisted} partition function $Z_{AB}$ since 
knowing 
its  character for all $g\in G$ helps us
 identify the nature of the associated representation. 

To be more precise: We are interested in knowing whether 
the representation 
${\hat g}$ in Eq.\  \eqref{LabelEqghat}
is projective
or not. 
On the other hand, as will be explained in the next section, knowing only
the values of character of a representation
for all $g\in G$, one cannot determine whether the representation 
is projective. 
This is only possible once we know the two-cocycle.
Therefore, coming back to the context of SPT phases: In  order to 
diagnose 
whether a projective representation 
occurs
in the spectrum or not,
i.e., in order to diagnose 
whether the boundary states $A$ and $B$ correspond to topologically distinct gapped phases,
we propose a diagnostic 
that we call the {\it symmetry-enforced vanishing of the partition function}, to be discussed in the next section.

\subsection{Symmetry-enforced vanishing of the partition function} 
\label{LabelSectionSymmetryEnforcedVanishing}

To illustrate the notion of the symmetry-enforced vanishing of the partition function,
which we will define momentarily in a more precise fashion,
we note the following properties of projective representations of a discrete group $G$.
First of all, it is well known\footnote{See e.g. Ref. [\onlinecite{M-Isaacs-CharacterTheory}].} 
 that the character of a {\it non-projective} irreducible representation of a finite group
always vanishes on at least one group element unless the representation is 
one-dimensional.
We will refer to this as an {\it accidental vanishing}
of the character.
On the other hand, the
character of a {\it projective}  representation, irreducible or not,  is {\it forced} to vanish
in the following sense \cite{Billo2001}:
for a given projective representation
with two-cocycle $\omega(g|h)$,
we define
\begin{align}
\varepsilon(g|h) = \omega(g|h) \omega(h|g)^{-1}, \quad {\rm when} \ \   [g,h]=e. 
\label{discrete torsion and two-cocyles}
\end{align}
The 
character $\rho$  of the projective representation evaluated on the group element $h$ vanishes, $\rho(h)=0$, 
if a group element $g\in N_h$
exists such that $\varepsilon(g|h)\neq 1$. 
To see this, we note that  $\rho(ghg^{-1})$ can be written as, 
\begin{align}
\rho(ghg^{-1}) &= 
\omega(gh|g^{-1})^{-1} 
\omega(g|h)^{-1} 
\omega(g^{-1}|g)
\nonumber \\
&\quad \times
\omega (h|e) \rho(h).
\label{useful id}
\end{align}
In particular, when $g$ and $h$ commute,  
$
\rho(h) = \varepsilon(h|g)\rho(h)
$
and hence $\rho(h)$ must vanish\footnote{Clearly, $\varepsilon(g|h)\neq 1$ also implies $\rho(g)=0$ by the same argument.}
when 
$\varepsilon(g|h)\neq 1$.
\footnote{We note that  the representation $\rho$ is {\it not} projective
if and only if
$\varepsilon(g|h)= 1$ for all commuting group elements $g$ and $h$. Moreover, as already mentioned in the paragraph above
Eq.\ (\ref{discrete torsion and two-cocyles}),
the dimension of a representation must
be greater than one if its character vanishes on at least one of the group elements.}
Thus in short, while the vanishing of its character alone does not
allow us to 
determine whether  the representation is projective or non-projective, 
if the vanishing is enforced, in the above sense,
this gives us a strong indication that the corresponding representation is projective. 

Similar to the above statement at the level of the group character $\rho(h)$,
we will argue below that the orbifold partition function 
allows us to infer whether non-trivial  two-cocyles of the representations are  included in the partition sum,
i.e. whether the representations are projective.
In particular, we introduce the notion of the {\it symmetry-enforced vanishing of the partition function.}
Consider the case where the twisted partition function vanishes,   
\begin{align}
Z^h_{AB}=\mathrm{Tr}_{\mathcal{H}_{AB} }\, \left[ \hat{h} \ e^{-\beta \hat{H}^{open}_{AB}} \right]=0,
\end{align}
and where  this vanishing of $Z^h_{AB}$ is enforced by symmetry. 
(Note that in view of Eq.\ (\ref{TwistedOpenStringPartitionFunction})
this vanishing  implies that  $\rho_a(h)=0$ for all irreducible representations $a$  occurring in Eq.\ (\ref{decompose}), 
since the conformal characters $\chi_a(q)$ are linearly independent.
Then, since
the character of a one-dimensional representation does not vanish, this  implies in view of
Eqs.\  \eqref{UntwistedOpenStringPartitionFunction} and \eqref{RhoOneEqualsMultiplicity} 
the appearance of multiplicities
$n^a_{AB} > 1$ for all irreps  in the spectrum of 
${\hat H}^{open}_{AB}$.)
In order to formulate the precise meaning of 
the symmetry enforced vanishing, it is convenient to go to
the 
closed string  picture, in which $Z_{AB}^h$ can be expressed
in terms of boundary states
\begin{align}
Z^{h}_{AB}=
{ }^{\ }_h\langle A|
e^{ -\frac{\ell}{2} \hat{H}^{closed} }
|B\rangle^{\ }_h 
=
{ }^{\ }_{h}\langle A| 
\tilde{q}^{\frac{1}{2}
(\hat{H}_L+\hat{H}_R) } 
|B\rangle^{\ }_{h}, 
\end{align}
which generalizes Eq.\ \eqref{ZABPartitionFunctionClosedStringPicture}.
Here 
$|A\rangle_h$ is a boundary state 
in the sector twisted by the group element $h\in G$. 
Thus, the vanishing of $Z^h_{AB}$ means 
\begin{align}
{ }^{\ }_h\langle A|
e^{ -\frac{\ell}{2} \hat{H}^{closed} }
|B\rangle^{\ }_h 
=0.
\end{align}

Suppose now the 
boundary state in the sector twisted by $h$ 
is not invariant under the
 symmetry\footnote{Here ${\hat g}$ denotes the representation of the group element $g$ on the Hilbert space of the
bulk CFT, of which the boundary state $|B>$ and its twisted variant $|B>_h$ are elements.}
operation
$g$, but picks up an {\it ``anomalous  phase''}\footnote{Note that the so-defined phase factor $\varepsilon_B(g|h)$ is an object
{\it entirely different} from the phase $\varepsilon(g|h)$ - no subscript $B$- defined in Eq.\ (\ref{discrete torsion and two-cocyles}).
}
factor $\varepsilon_B(g|h)$: 
\begin{align}
\hat{g} | B\rangle_{h} = \varepsilon_B(g|h) |B\rangle_{h},
\quad
\mbox{when} \quad
g\in N_{h}.
\label{anomalous phase}
\end{align}
Then, since $\hat{g}$ is a symmetry of the Hamiltonian,  we have
\begin{align}
&
{ }^{\ }_{h}\langle A| \hat{g} \tilde{q}^{\frac{1}{2}(\hat{H}_L+\hat{H}_R) } 
|B\rangle^{\ }_{h} 
= 
{ }^{\ }_{h} \langle A| \tilde{q}^{\frac{1}{2} (\hat{H}_L+\hat{H}_R) } \hat{g}
|B\rangle^{\ }_{h}, 
\end{align}
from which 
it follows that
\begin{align}\label{VanishingParitionFunctionAnomalousPhase}
&
\varepsilon_A(g|h)^*
{ }^{\ }_{h}
\langle A|  \tilde{q}^{\frac{1}{2} (\hat{H}_L+\hat{H}_R) } |B\rangle^{\ }_{h}
\nonumber \\
&
\quad
= 
\varepsilon_B(g|h)
{ }^{\ }_{h}\langle A| \tilde{q}^{\frac{\ell}{2} (\hat{H}_L+\hat{H}_R) }  |B\rangle^{\ }_{h}.
\end{align}
Thus, unless 
$\varepsilon_A(g|h)^* = \varepsilon_B(g|h)$, 
the twisted partition function must vanish.  
When the partition function vanishes due to the anomalous phases
$\varepsilon_A(g|h)$ and $\varepsilon_B(g|h)$,
we call this situation
a {\it symmetry-enforced vanishing} of the (twisted) partition function. 
Note that
Eq.\ \eqref{VanishingParitionFunctionAnomalousPhase} reads, 
in view of Eq.\ \eqref{TwistedOpenStringPartitionFunction},
\begin{equation}
\rho_a(h) = { \varepsilon_B(g|h) \over \varepsilon^*_A(g|h)} \ \rho_a(h), \qquad g \in N_h
\end{equation}
for all irreducible representations $a$ appearing in
Eq.\ \eqref{decompose}.
Therefore, we argue that when this happens, 
the gapped phase
which sets 
the corresponding boundary condition, and which hence determines the boundary state,
is a non-trivial SPT phase.

In the following sections,  we will demonstrate that such  a symmetry-enforced vanishing
of the twisted partition function occurs indeed  in 
various characteristic 
examples of SPT phases: the time-reversal breaking Kitaev (superconducting) chain in symmetry class D, the Haldane phase,
and the time-reversal invariant Majorana chain in symmetry class BDI.
Observe that the partition function may vanish {\it accidentally}
even when there is no anomalous phase. 
This should be distinguished from the vanishing of the partition function
which is enforced by symmetry. 
In general, we do not expect  a vanishing of the partition function  
which is not enforced is a consequence of the topological features of an SPT phase. 

The assumption we made in Eq.\ \eqref{anomalous phase} deserves more discussion,
since there are in principle more generic possibilities
for the action of the symmetry on boundary states, besides the one listed in
Eq.\ \eqref{anomalous phase}.  
When boundary conditions (boundary states)
break symmetries, 
we expect the symmetry operation $\hat{g}$ will in general 
map  one boundary state into another. 
On the other hand, for boundary states that arise from  (1+1) dimensional SPT phases, we do not
expect that they break the symmetry defining the SPT phase. Hence,  
one may expect that  the symmetry operation leaves boundary states invariant (up to a phase),
as in Eq.\ \eqref{anomalous phase}. 
(This point will be further illustrated in the next section). 
However, in principle, there is a logical possibility that $\hat{g}$ 
maps 
a boundary state into another 
 boundary state.
I.e., there
could in principle exist a multiplet of boundary states 
that are mapped on to each other by $\hat{g}$.
While we do not have a formal proof, in  all examples of SPTs
we looked at, a given boundary states is a  singlet under the symmetry defining the SPT phase,
as in Eq.\ \eqref{anomalous phase}. 
However, for other more complicated examples, 
there may be a multiplet of boundary states.

Let us contrast this with a 
slightly different context.
In Ref.\ \onlinecite{Hanunpublished},
(1+1)d CFTs
which appear at boundaries (edges) of (2+1)d SPT phases
are considered.
These (1+1)d edge theories of (2+1)d SPT phases
are expected to be 'ingappable' once
the symmetry defining 
the SPTs are strictly enforced (also on the (1+1)d edge theory) . 
In Ref.\ \onlinecite{Hanunpublished},
possible boundary conditions (boundary states) 
in the (1+1)d  edge theories are investigated 
for various examples of (2+1)d SPT phases.
It is found that there exist   no conformally invariant boundary
conditions on these (1+1)d CFTs that preserve the symmetries of the underlying 
(2+1)d SPT phase. In other words, all conformally invariant boundary conditions
of these (1+1)d CFTs are not invariant under the symmetry.
This is expected since as boundaries 
of the (2+1)d SPTs, these (1+1)d CFT cannot themselves have boundaries,
and since this statement refers to such theories that respect the symmetry of the SPT phase.
Indeed, viewed from the perspective of the present paper, if a (1+1)d CFT appearing at the boundary
of the (2+1)d SPT was gappable while preserving all the symmetries of the SPT, then the entanglement
spectrum of that gapped  (1+1)d theory at the boundary would be a BCFT with boundary conditions preserving the
symmetries of the SPT. Thus the absence of boundary conditions on the (1+1)d CFT which preserve the symmetries of the SPT, 
implies that this CFT is ingappable.


\subsection{Boundary conditions and anomalous phases} 
\label{Boundary conditions and anomalous phases} 

In order to  provide  more intuition 
about  the symmetry-enforced vanishing of the partition function,
let us now show that the anomalous phase 
\eqref{anomalous phase}
can be interpreted as a quantum anomaly.
As we discussed in the previous sections,
we associate a BCFT 
with  a SPT phase;
the SPT serves as a boundary condition 
on a given CFT. 
By rotating (Euclidean) spacetime by $\pi/2$, namely
$(x,\tau)=$ $  (-{\tilde \tau}, {\tilde x})=:$
 $(\sigma_1, \sigma_2)$, 
we then introduce boundary states located
at an ``initial'' imaginary time in the rotated coordinates, ${\tilde \tau} = \sigma_1 =0$,
in the form
\begin{align}
\label{LabelEqBoundaryConditionsOnFields}
&
\left[
\hat{\Phi}(\sigma_2)
-
U \hat{\Phi}(\sigma_2)
\right]|B\rangle_h
=0, 
\qquad ({\rm at}  \ \sigma_1=0)
\end{align}
which encode  the boundary condition located at $x=0$ in the unrotated coordinates.
Here, 
$\hat{\Phi}(\sigma_2)$ denotes  a  (column) vector of quantum field operators representing fundamental degrees of freedom
of the CFT under consideration, 
$U$ is a matrix acting on the 
column vector  ${\hat \Phi}$ of fields, 
and 
$|\cdots\rangle_h$ represents a state in the $h$-twisted sector.
By definition, 
states in the $h$-twisted sector obey
\begin{align}
 \left[\hat{\Phi}(\sigma_2+\beta) - \hat{h} 
 \hat{\Phi}(\sigma_2)\hat{h}^{-1} \right] |\cdots \rangle_h =0.
\end{align}
Note that for a given boundary state,
there may not be a simple description in terms of a fundamental field $\hat{\Phi}$,
as that given in Eq.\ \eqref{LabelEqBoundaryConditionsOnFields}.
However, when such description is available, we can develop an intuitive picture as follows.

Let the symmetry $g$ act on fundamental fields $\hat{\Phi}$ as
\begin{align}
\hat{g}
\hat{\Phi}(\sigma_2)
\hat{g}^{-1}
=
U_{g} \hat{\Phi}(\sigma_2), 
\end{align}
where $U_g$ is a matrix acting on the components of the (column) vector  $\hat{\Phi}$.  Let us now
act with $g$ on the boundary condition, 
\begin{align}
\label{LabelEqGroupInvarianceOfBoundaryCondition}
 &\quad 
 \left[
 \hat{\Phi}(\sigma_2 ) 
- 
U \hat{\Phi}(\sigma_2) 
\right]
| B \rangle_h 
=0
\nonumber \\
 &\Rightarrow 
 \hat{g}
 \left[
 \hat{\Phi}(\sigma_2) 
- 
U \hat{\Phi}(\sigma_2) 
\right]
\hat{g}^{-1}
\hat{g}
| B \rangle_{h}
=0
\nonumber \\
 &\Rightarrow 
 \left[
  U_{{g}} \hat{\Phi}(\sigma_2) 
- 
U U_{{g}} \hat{\Phi}(\sigma_2) 
\right]
\hat{g}
| B\rangle_h
=0
\nonumber \\
 &\Rightarrow 
 \left[
 \hat{\Phi}(\sigma_2 ) 
- 
U^{-1}_{{g}} 
U^{\ } 
U^{\ }_{{g}} \hat{\Phi}(\sigma_2) 
\right]
\hat{g}|B\rangle_h 
=0. 
\end{align}
By definition, 
our problem preserves the symmetry $g$, and hence we should have
$U^{-1}_{g} U U_{g} =U$. 
If the boundary condition is invariant,
then we may expect that so is the boundary state, 
$\hat{g}|B \rangle_h = |B \rangle_h$.
However this expected invariance may be broken  
quantum mechanically;
the boundary state may not be invariant, but may acquire a phase, 
$\varepsilon_B(g|h)$,
under the action of the symmetry.
The phase $\varepsilon_B(g|h)$ can then be considered as a kind 
of quantum anomaly.
While the boundary condition is invariant under the  symmetry, 
the corresponding quantum mechanical state may not be.
This anomaly signals 
the non-trivial topological properties of the corresponding ``bulk" SPT phase.

\section{The Kitaev chain (Class D)}
\label{The Kitaev chain}

In this section, we apply the discussion from  the preceding section
to a simple fermionic SPT phase in (1+1)d,
the Kitaev chain. 
The Kitaev chain is a fermionic SPT phase protected by 
fermion number parity conservation 
(symmetry class D).

In the continuum limit the Kitaev chain is described by the action \eqref{Kitaev chain},
or equivalently in terms of the Hamiltonian 
\begin{align}
H&=H_0+ H_I, 
\nonumber \\
H_0
&=
\int^{\ell}_0dx\, 
\left[
\psi_L ( +v i\partial_x) \psi_L
+
\psi_R ( - v i \partial_x) \psi_R
\right],
\nonumber \\
H_I 
&=
\int^{\ell}_0 dx\, im \psi_L \psi_R, 
\label{fermionic edge theory single flavor}
\end{align}
where (anti-)periodic boundary  conditions on the Majorana fermions are imposed,
$\psi_L(x+l)=$ $\pm \psi_L(x)$, $\psi_R(x+l)=$ $\pm \psi_R(x)$.
The fermi velocity $v$ was set to unity for simplicity. 
The real fermion fields $\psi_L, \psi_R$ obey the canonical anticommutation relations
\begin{align}
\{ \psi_L(x), \psi_L(x')\} &= 2\pi \sum_{n\in \mathbb{Z}} \delta(x-x'+\ell n),
\nonumber \\
\{ \psi_R(x), \psi_R(x')\} &= 2\pi \sum_{n\in \mathbb{Z}} \delta(x-x'+\ell n).
\end{align}
The fermionic Hamiltonian
 (\ref{fermionic edge theory single flavor}) preserves fermion number parity,
$[H, \hat{g}_f]=0$, where 
\begin{align}
\hat{g}_f = 
(-1)^{F},
\quad
 F = \frac{1}{2\pi} \int^{\ell}_0dx\,  i \psi_L \psi_R. 
\end{align}
Fermion parity ${\hat g}_f$ is the {\it only} symmetry of the Hamiltonian that we consider in this section (which
is a member of symmetry class D). I.e., the symmetry group protecting the SPT in Eq. \ (\ref{fermionic edge theory single flavor})
(and its $N_f$-flavor generalization discussed below)  is $G=\mathbb{Z}^F_2$ where the superscript ${}^F$ stands for fermion parity.

We also consider the generalization to $N_f$ flavors of real (Majorana) fermions
described by the Hamiltonian 
\begin{align}
H
&=
\sum^{N_f}_{a=1}
\int^{\ell}_0dx\, 
\left[
\psi^a_L ( + i\partial_x) \psi^a_L
+
\psi^a_R ( -  i \partial_x) \psi^a_R
\right]. 
\label{fermionic edge theory}
\end{align}
The fermion fields obey the canonical anticommutation relations
\begin{align}
\{ \psi^a_L(x), \psi^b_L(x')\} &= 2\pi \delta^{ab} \sum_{m\in \mathbb{Z}} \delta(x-x'+\ell m),
\nonumber \\
\{ \psi^a_R(x), \psi^b_R(x')\} &= 2\pi \delta^{ab} \sum_{m\in \mathbb{Z}} \delta(x-x'+\ell m).
\end{align}
The Hamiltonian of the fermionic theory  with $N_f$ flavors  in Eq.\ (\ref{fermionic edge theory})
commutes with the total fermion number parity operator,
given by
\begin{align}
\label{LabelDEFTotalFermionParitityNumberOperator}
\hat{g}_f = 
(-1)^{F},
\quad
F =\sum^{N_f}_{a=1} F_a, 
\end{align}
where $F_a$ is the total fermion number operator for the $a$-th flavor, 
\begin{align}
\label{LabelDEF-FermionParitya}
 F_a = \frac{1}{2\pi} \int^{\ell}_0dx\,  i \psi^a_L \psi^a_R. 
\end{align}

The Hamiltonian \eqref{fermionic edge theory single flavor}
realizes two gapped phases separated by a
quantum phase transition at $m=0$. 
The two gapped phases 
can be topologically distinguished by a $\mathbb{Z}_2$ topological invariant. 
(Which sign of the mass term realizes the  topological or trivial phase 
cannot be distinguished from the above continuum model, 
but the relative topological charge of the two gapped phases is well-defined.)
It is well known that the entanglement spectrum of the topologically non-trivial phase 
is at least two-fold degenerate,
while 
that of the trivial phase does not support any degeneracy.
\cite{Ryu2006, turner2011topological,ChoLudwigRyuarXiv2016}

Following the 
discussion in Sec.\ \ref{LabelSectionEntanglementSpectrum},
the low-lying  entanglement spectrum is described in the scaling limit  by the spectrum of an appropriate BCFT,
i.e., an appropriate CFT with the boundary conditions specified by the topological properties of the gapped SPT phase.  
The spectrum of the BCFT is described (upon folding) by 
a chiral CFT defined on a circle of length $2\times \ell$,  
\begin{align}
 H = \int_0^{2\ell}  dx\, \psi_L i \partial_x \psi_L,
\end{align}
where
the fermion field obeys either 
the antiperiodic ('NS') or the periodic  ('R') boundary conditions,
\begin{align}
 \psi_L(x+2\ell) = -\psi_L(x),
 \quad
 \mbox{or}
 \quad
 \psi_L(x+2\ell) = +\psi_L(x).
\end{align}
These two boundary conditions, i.e., two different BCFTs, 
correspond to the trivial and topological states of the Kitaev chain \eqref{Kitaev chain},
as we will review  momentarily. 

Corresponding to these two  boundary conditions,
we consider the partition functions
\begin{align}
\label{PartitionFctsMajorana-AA-PA}
{\tilde Z}_{AA}(q) = \mathrm{Tr}_{A} \ q^{H_L},
\quad
{\tilde Z}_{PA}(q) = \mathrm{Tr}_{P} \ q^{H_L}.
\end{align}
Here, ${\tilde Z}_{AB}$ denotes   the chiral partition function with spatial and temporal periodicity conditions
labeled by $A$ and $B$, respectively;
and $P$ ($A$) stand for  periodic (antiperiodic) boundary 
conditions.\footnote{Note that ${\tilde Z}_{AB}$ is technically an object different from $Z_{AB}$ in
Eq.\ (\ref{OpenStringPartitionFunctionUnprojected}), since in the latter the subscripts denote boundary conditions
on a non-chiral CFT defined on a finite interval, whereas in the former the subscripts denote periodicity conditions
on the chiral (say, only left-moving) fermion degrees of freedom (even there is of course a connection, which is recalled below).
For that reason the former partition function is distinguished from the latter by a different symbol.}
In Eq.\ (\ref{PartitionFctsMajorana-AA-PA})
the temporal direction is always anti-periodic (which is well known to follow in general from the Fermion path integral).
In addition to these partition functions 
we consider,
following our discussion in Sec. \ref{LabelSection-SymmetryProtectedDegeneracyEntanglementSpectrumGeneralConsiderations},
 the sector twisted by the only non-trivial group element of the symmetry group,  the fermion number parity operator.
(Recall that fermion number parity is the only symmetry of the Hamiltonian in symmetry class D,
which we consider in this section.)
We are thus lead to consider, in addition to Eq.\ (\ref{PartitionFctsMajorana-AA-PA}), the partition functions
\begin{align}
{\tilde Z}_{AP}(q) = \mathrm{Tr}_{A} (-1)^F q^{H_L},
\quad
{\tilde Z}_{PP}(q)= \mathrm{Tr}_{P} (-1)^F q^{H_L}.
\end{align}

As  is well-known, ${\tilde Z}_{PP}$ actually vanishes, ${\tilde Z}_{PP}=0$.
This  is due to the fermion zero mode.
(This should be distinguished from the 
zero mode that causes the symmetry protected degeneracy in the
entanglement spectrum we are after.) 
As we will now explain, the vanishing  ${\tilde Z}_{PP}=0$
 is precisely an example of 
a symmetry-enforced vanishing of the partition function discussed
in general terms
 in Sec. \ref{LabelSection-SymmetryProtectedDegeneracyEntanglementSpectrumGeneralConsiderations}, the
symmetry being Fermion number  parity. 
As we will explain in the following, when this partition function is described within   
the boundary state formalism the corresponding boundary state will
pick up an anomalous phase.

\subsection{Boundary states}

To discuss the symmetry-enforced vanishing of the partition function of the current theory (class D)
 from  the CFT point of view, we consider the free fermion CFT that results from
setting $m=0$
in the Hamiltonian \eqref{Kitaev chain}. 
Consider this (gapless) free-fermion CFT on the interval $x_1 \leq x \leq  x_2$. 
At the two boundaries  $x=x_1$ and $x=x_2$ of this interval, let us
consider the following boundary conditions  on the fermion 
field\footnote{The difference in sign for the boundary conditions
at $x=x_1$ and $x=x_2$ arises from the fact that both $\psi_L(z)$ and $\psi_R(\bar z)$
transform as spinors under rotations in Euclidean two-dimensional spacetime, where $z = \tau + i x$,
${\bar z} = \tau - i x$ (or, equivalently, from the fact that they  have conformal
weight (`scaling dimension') $h=1/2$). That  this leads to the signs displayed in Eq.\ (\ref{LabelEqFreeFermionBoundaryConditions})
can easily be seen as follows. Consider first the situation where $x_1=0$ and $x_2 \to +\infty$; there is hence
only one boundary, namely the one at $x=x_1 \to 0$. The spacetime
is then the upper half complex plane, $\mathrm{Im} \  z \geq 0$, and the boundary is located on the real axis, $\mathrm{Im} \  z = 0$,
which is a ``lower boundary''. Compare
this with the situation where $x_2=0$ and $x_1 \to -\infty$; there is now also only  one boundary, namely the one
at $x_2=0$. The spacetime is now the lower half  complex plane, $\mathrm{Im} \ z \leq 0$, and the boundary is again located
on the real axis, which is however now an ``upper boundary''. - Now, the two situations of an ``upper boundary'' and of a ``lower boundary''
are related to each other by reflection about the real axis, $z \to -z$, and ${\bar z} \to -{\bar z}$. Let us impose
on the ``lower boundary'' the condition
$\psi_L(x_1=0)=\eta_1 \psi_R(x_1=0)$, or equivalently  $\psi_L(z)=\eta_1 \psi_R({\bar z})$ when $z = {\bar z}$ (hence ${Im} \  z =0$).
Then the {\it same} boundary condition would read at an ``upper boundary''  $e^{+i \pi/2} \psi_L(-z)=\eta_1  e^{-i \pi/2} \psi_R(-{\bar z})$, implying  $\psi_L(-z)=-\eta_1  \psi_R(-{\bar z})$
when $z={\bar z}$ (thus ${Im} \ z =0$), and therefore $\psi_L(x_2=0)= - \eta_1 \psi_R(x_2=0)$. [Here we used $\psi_L(e^{i\alpha} z) = (1/e^{+i\alpha/2})\psi_L(z)$,
and $\psi_R(e^{i\alpha} {\bar z}) = (1/e^{-i\alpha/2})\psi_R({\bar z})$.]
For this reason,
the {\it same} boundary condition appears with the opposite sign of $\eta_1$ at the ``upper boundary' as compared to the
``lower boundary'''.}

\begin{align}
\label{LabelEqFreeFermionBoundaryConditions}
\psi_L(x_1) =\eta_1 \psi_R(x_1), 
\quad
\psi_L(x_2) = -\eta_2 \psi_R(x_2),
\end{align}
where $\eta_{1}, \eta_{2}=\pm 1$.
In terms of the scattering matrix language discussed in Sec.\ \ref{LabelSubSection-ScatteringFromTopPhases},
these  boundary conditions correspond to
the reflection coefficients (matrices) 
\begin{align}
r = \eta_2, \quad r' = \eta_1.
\end{align}
(Compare Eq.\ (\ref{LabelEq-DEF-ScatteringMatrix}).)
The topological invariant computed from these reflection coefficients (matrices)  is given by
\begin{align}
 \mathrm{sgn}\, \det r = \eta_2,
\qquad
 \mathrm{sgn}\, \det r = \eta_1.
\end{align}
When $\eta_1=\eta_2$ we obtain  (upon employing the `folding procedure') from 
Eq.\ (\ref{LabelEqFreeFermionBoundaryConditions}) a system of chiral (say, left-moving)
fermions on an interval of length $ 2\ell$ with {\it anti-periodic} (`NS')  boundary conditions.
As is well known, 
this spectrum has no degeneracies. On the other hand, when $\eta_1=-\eta_2$,
the resulting system of chiral (say, left-moving)
fermions on an interval of length $2\ell$ has {\it periodic} (``R'') boundary condtions,
which has a two-fold degeneracy (as is also well known).
The choice
 $\eta_1 =- \eta_2$ for the pair of boundary conditions corresponds to   a domain wall in the mass term - see Fig.\ \ref{Fig1}. 
Thus the condition $\eta_1 =- \eta_2$ localizes a non-trivial zero mode  in the
gapless region $x_1 \leq x \leq x_2$  by the  Jackiw-Rebbi mechanism,
as discussed in Sec. \ \ref{LabelSubSection-ScatteringFromTopPhases}.

Let us now construct the 
boundary states corresponding to the boundary conditions in Eq. (\ref{LabelEqFreeFermionBoundaryConditions}).
 Consider e.g. the boundary condition at $x=x_1$, given by
 \begin{align}
\label{LabelEqBoundaryConditionFermionsx1}
\psi_L(\tau, x_1) &=\eta_1 \psi_R(\tau, x_1), \quad (0\leq \tau \leq \beta),
\end{align}
where the fermions $\psi_L$ and $\psi_R$ possess their (natural)  anti-periodic boundary conditions in imaginary
time $\tau$.
We now make the 
rotation by $\pi/2$ of  (Euclidean) spacetime discussed in the paragraph 
surrounding Eq. \ref{LabelEqBoundaryConditionsOnFields}, namely $(x,\tau)=$ $  (-{\tilde \tau}, {\tilde x})$.
Since the fermion fields $\psi_L$ ($\psi_R$) are holomorphic (anti-holomorphic) functions of conformal weight (scaling dimension)
$1/2$, they transform
under the $\pi/2$ rotation $(\tau+ix)=$ $({\tilde x}-i {\tilde \tau})=$ $(-i) ({\tilde \tau} + i {\tilde x})$
and  $(\tau-ix)=$  $(+i) ({\tilde \tau} - i {\tilde x})$
as
\begin{align}
\label{LabelEqClassDPiOverTwoRotation}
\psi_L(\tau+ix) =e^{+i\pi/4}\psi_L({\tilde \tau} + i {\tilde x}),
\\ \nonumber
\psi_R(\tau-ix) =e^{-i\pi/4}\psi_R({\tilde \tau} - i {\tilde x}).
\end{align}
This implies that the boundary condition (\ref{LabelEqBoundaryConditionFermionsx1}) reads in the rotated coordinates
\begin{align}
\label{LabelEqBoundaryConditionFermionsx1rotated}
\psi_L({\tilde \tau}_1,{\tilde  x}) =(-i) \eta_1 \psi_R({\tilde \tau}_1, {\tilde x}), \quad (0\leq {\tilde x}  \leq \beta).
\end{align}
The boundary states $|B(\eta)\rangle$  represent an operator statement
of the boundary condition (\ref{LabelEqBoundaryConditionFermionsx1rotated})
on the closed string Hilbert space,
\begin{align}
\label{LabelEqDEFBoundaryStates}
  \left[
   \psi_L ({\tilde \tau}, {\tilde x})- i (-\eta_1) \psi_R ({\tilde \tau}, {\tilde x})
  \right]
  |B (-\eta_1) \rangle &=
  0, 
\end{align}
with {\it anti-periodicity} in  $0 \leq {\tilde x} \leq \beta$, which is inherited from the anti-periodicity in $\tau$.
 For simplicity we now set ${\tilde \tau} = x_1=0$,
and omit writing  the ${\tilde \tau}$ coordinate. The boundary state describing the boundary condition at $x=x_2$ satisfies
the same equation with $\eta_1 \to \eta_2$  ({\it not}
 $\eta_1 \to - \eta_2$;
see the footnote immediately above Eq.\ (\ref{LabelEqFreeFermionBoundaryConditions})).

Following the discussion in 
Sec.\ \ref{LabelSection-SymmetryProtectedDegeneracyEntanglementSpectrumGeneralConsiderations}
we must now twist the boundary states defined in Eq.\ (\ref{LabelEqDEFBoundaryStates}) by a group element of the symmetry
group of the SPT phase. As mentioned above, in the current case of symmetry class D, there is only one non-trivial
group element, which is the fermion parity operator $\hat{g}_f$ defined in
Eq.\ (\ref{LabelDEFTotalFermionParitityNumberOperator}).
Because the fermion parity operators changes the periodicity on both, the left- and the right-moving fermions
in Eq.\ (\ref{LabelEqDEFBoundaryStates}) from anti-periodic to periodic, the twisted boundary state $|B(\eta)\rangle_{\hat{g}_f}$
satisfies the equation
\begin{align}
\label{LabelEqDEFBoundaryStatesTwisted}
  \left[
   \psi_L ({\tilde x})- i \eta \psi_R ({\tilde x})
  \right] 
  |B (\eta) \rangle_{\hat{g}_f} &=
  0, 
 \end{align}
with {\it periodicity} in  $0 \leq {\tilde x} \leq \beta$.
Upon Fourier transforming the Hermitean (Majorana) fermion operators,
\begin{align}
\label{LabelEqFourierFermionOperators}
\psi_L({\tilde x}) ={2\pi\over \beta} \sum_s e^{-i 2\pi {\tilde x} s/\beta} \psi_{sL},
\quad  \psi^\dagger_{s L}=  \psi_{-s L}
\\ 
\psi_R({\tilde x}) ={2\pi\over \beta} \sum_s e^{+i 2\pi {\tilde x} s/\beta} \psi_{sR},
\quad  \psi^\dagger_{s R}=  \psi_{-s R}
\end{align}
where the mode-index $s\in \mathbb{Z}+{1\over 2}$ ($s\in \mathbb{Z}$) for anti-periodic (periodic) boundary conditions in ${\tilde x}$,
Eq.\ (\ref{LabelEqDEFBoundaryStatesTwisted}) reads
\begin{align}
\label{LabelEqDEFTwistedBoundaryStateMomentum}
[\psi_{s L} -i \eta \psi^\dagger_{s R}] |B(\eta)\rangle_{\hat{g}_f}=0, \qquad  (s \in \mathbb{Z}).
\end{align}
This determines the boundary state to be of the form
\begin{align}
\label{LabelEqExplicitFormBoundaryState}
 |B(\eta)\rangle_{\hat{g}_f}
=
\exp\left\{
{i\over (-\eta)}  \sum_{s=1}^{+\infty}\psi^\dagger_{s R} \psi^\dagger_{s L} 
\right\} \   
|B(\eta)\rangle_0,  
\end{align}
where the ``zero-mode contribution'' (from $s=0$),  $|B(\eta)\rangle_0$, is determined by
\begin{align}
\label{LabelEqZeroModeContributionBoundaryState}
  \left[
   \psi_{0L} - i \eta \psi_{0R} 
  \right]
  |B, \eta\rangle_0
  &=
  0.
\end{align}
The zero modes satisfy $(\psi_{0L})^2=$ $(\psi_{0R})^2=1$.
The zero-mode contribution to the boundary state 
can be constructed 
by considering
the following fermion creation and annihilation operators (we immediately
discuss here  the general case of  $N_f$ Majorana flavors, $a=1, ..., N_f$), 
\begin{align}
f^{\dag}_{a} &= 
\frac{1}{{2}}(\psi^a_{0L} + {i}\psi^a_{0R}), 
\quad
f^{\ }_{a} =
\frac{1}{{2}}(\psi^a_{0L} - {i}\psi^a_{0R}),
\end{align}
where $|0_f\rangle$  denotes
the Fock vacuum of the $f_a$-fermions. 
In view of Eqs. (\ref{LabelEqDEFTwistedBoundaryStateMomentum}, \ref{LabelEqFourierFermionOperators}),
the boundary state $|B, \eta=+\rangle_0$ is then nothing
but the Fock vacuum
$|0_f\rangle$ itself, 
\begin{align}
 |B, \eta =+\rangle_0 =  e^{ i\phi_+}|0_f\rangle.  
\end{align}
On the other hand,
the boundary state $|B, \eta=-\rangle_0$ can be constructed as
\begin{align}
 |B, \eta =-\rangle_0 = e^{i\phi_-} \prod_{a=1}^{N_f} f^{\dag}_a  |0_f\rangle. 
\end{align}

In the above representation of $|B,\eta\rangle_0$,
the ambiguous phases $\phi_{\pm}$ are not fixed by the boundary condition.
These phases will not affect our analysis in this section, and hence will be set to zero henceforth. 
We will come back to the issue of a suitable 
choice of the phase at the end of this section,
and also in Sec.\ \ref{BDI topological superconductors in (1+1)d}, in which 
a  proper choice of the phase is more crucial.
[We however comment that one common convention for the phase is 
$
|B, \eta = + \rangle_0 = e^{-i\frac{\pi}{8}}|0_f\rangle 
$, 
$
|B, \eta = - \rangle_0 = e^{i\frac{\pi}{8}} f^{\dag}|0_f \rangle.
$
(These equations are written, for simplicity, for the case of a  single flavor $N_f=1$.)
One motivation for this phase convention is that 
the following relations 
$
\psi_{0L} |B, \eta \rangle_0 = e^{-i\eta \frac{\pi}{4}}|B, -\eta \rangle 
$, 
$
\psi_{0R} |B, \eta \rangle_0 = e^{-i\eta \frac{\pi}{4}}|B, -\eta \rangle
$
bear resemblance 
to the operator product expansions of the Ising CFT, 
$\psi_{L} \sigma \sim e^{i\frac{\pi}{4}}\mu$,
$\psi_{L} \mu \sim  e^{-i \frac{\pi}{4}}\sigma$,
$\psi_{R} \sigma \sim e^{-i\frac{\pi}{4}} \mu$, 
$\psi_{R} \mu \sim e^{i \frac{\pi}{4}} \sigma$,
where $\sigma$ and $\mu$ are the Ising spin operator and the disorder operator, respectively.
(See e.g. Appendix E of  Ref.\ [\onlinecite{BPZ}], or Ref.\ [\onlinecite{CFTbook}].)]

Following the general discussion of Sec. \ref{LabelSectionSymmetryEnforcedVanishing}, in particular  Eq. (\ref{anomalous phase}),
we now ask about  the properties of the boundary states 
under the action of the fermion number parity $ \hat{g}_f $, the only element of the symmetry group of the present SPT.
Its explicit form within the zero mode sector of the closed string 
Hilbert space\footnote{The explicit form in terms of non-zero modes follows immediately
from Eq.\ (\ref{LabelDEFTotalFermionParitityNumberOperator},\ref{LabelDEF-FermionParitya},\ref{LabelEqFourierFermionOperators})}
 is
\begin{align}
 \hat{g}_f 
 =
 (i \psi^1_{0L}\psi^1_{0R})
 (i \psi^2_{0L}\psi^2_{0R})
 \cdots
 (i \psi^{N_f}_{0L}\psi^{N_f}_{0R}),
\end{align}
when there are the 
$N_f$ flavors of Majorana fermions.
This implies that the fermion number parity operator acting on the boundary states gives
\begin{align}
\hat{g}_f|B, \pm \rangle_0 = (\pm 1)^{N_f} |B, \pm\rangle_0. 
\label{sign D+R+}
\end{align}
Therefore, 
when $N_f=\mbox{even}$, there is no anomaly neither
for
$|B, + \rangle_0$ 
nor for
$|B, - \rangle_0$. 
On the other hand,
when $N_f=\mbox{odd}$
one would conclude that
$|B, -\rangle_0$ is anomalous
while 
$|B, +\rangle_0$ is not.
This is consistent with the $\mathbb{Z}_2$ classification of 
(1+1)d topological superconductors in symmetry class D. 

Upon closer inspection however, Eq.\ (\ref{sign D+R+}) would look strange 
since the two states $|B, \pm \rangle_0$
should be treated on the equal footing.
In fact, it should be noted that there is a phase ambiguity
in defining the boundary states and the fermion number parity operator. 
In the above analysis, we implicitly made a particular choice where  
the fermion number parity of the ground state $|0_f\rangle$ is $+1$. 
In principle, one could assign a different fermion number parity eigenvalue,
e.g., by modifying the definition of the fermion number parity operator,
$\hat{g}_f\to -\hat{g}_f$.
Alternatively, instead of using $f^{\dag}_a, f_a$,
one could define $c_a:=f^{\dag}_a$ and $c^{\dag}_a:=f_a$, 
which leads to 
$|B, -\rangle_0 =|0_c\rangle$
and 
$|B,+\rangle_0 =\prod_a c^{\dag}_a|0_c\rangle$. 
In this convention,
one would then be led to claim 
$\hat{g}_f|B,-\rangle_0 = +|B,-\rangle_0$
while 
$\hat{g}_f|B,+\rangle_0 = (-1)^{N_f}|B,+\rangle_0$. 
Thus, there is some ambiguity when deducing
the fermion number parity eigenvalue.
Such an  ambiguity of the fermion number parity eigenvalue of the ground state,
however, does not affect our conclusion,
since, independent of the phase choice,  
when $N_f=\mbox{odd}$, 
we cannot make both $|B,+\rangle_0$ and
$|B,-\rangle_0$ anomaly-free.
In conclusion, our analysis of the anomalous phase of the boundary state (as defined
in Eq. (\ref{anomalous phase}) of Sec. \ref{LabelSectionSymmetryEnforcedVanishing})
leads to the (known) result that there is a $\mathbb{Z}_2$ classification for (1+1)d SPT phases
in symmetry class D.

\section{The Haldane phase and the compact boson theory}
\label{The Haldane phase and the compact boson theory}

The Haldane phase of the SU(2)  spin-1 quantum spin chain
is historically the first and the canonical example of
a one-dimensional symmetry-protected topological phase. 
%
The Haldane phase was shown to be a stable 
symmetry-protected topological phase if one of the following 
discrete symmetries is imposed: 
\cite{pollmann2012symmetry}
\paragraph*{(i) TRS} 
Time-reversal acts on a spin-1 operator as 
\begin{align}
T:\quad 
\hat{T} \boldsymbol{S} \hat{T}^{-1} = - \boldsymbol{S}, 
\quad 
\hat{T} {i} \hat{T}^{-1} = -{i}. 
\end{align}
Note that ${T}^2=+1$.

\paragraph*{(ii) the dihedral group of $\pi$-rotations
about $x,y$ and $z$ axes ($D_2$)}
 
Consider a $\pi$-rotation around a particular vector
in spin space: e.g.,
$\pi$-rotation around $z$-axis is 
\begin{align}
R^z_{\pi}:
\quad 
S^x \to - S^x,
\quad
S^y \to - S^y,
\quad
S^z \to + S^z.
\end{align}
Take any two of $R^{x,y,z}_{\pi}$.
The third transformation is given by the
product of other two. 
So, this is $\mathbb{Z}_2\times \mathbb{Z}_2$ symmetry. 

\paragraph*{(iii) link inversion}
This can be realized as 
(site inversion) + (translation).
The one-site translation is given by
$
\boldsymbol{S}_j \to \boldsymbol{S}_{j+1},
$
while the site parity transformation is 
$
\boldsymbol{S}_j \to \boldsymbol{S}_{-j}. 
$
If combined, the link inversion $L$ acts on the spin operator 
at site $j$ as 
$
L:
\boldsymbol{S}_j \to \boldsymbol{S}_{j+1} \to \boldsymbol{S}_{-j-1}. 
$
%

In the following, we will focus on the protection of the Haldane phase
by $\mathbb{Z}_2\times \mathbb{Z}_2$ (dihedral) symmetry.



\subsection{Field theory descriptions of the Haldane phase}

The Haldane phase is known to be adjacent to at least three CFTs
the compactified free boson $U(1)$ ($c=1$),
the $SU(2)_2$ Wess-Zumino-Witten theory ($c=3/2$),
and 
the $SU(3)_1$ Wess-Zumino-Witten theory ($c=2$).
In this section, we will focus on the $c=1$ CFT, and discuss 
its neighboring gapped phase; 
the Haldane phase (non-trivial SPT phase)
and 
the so-called large $D$-phase. 

%
%

We start from the free boson theory on a spatial ring of circumference $\ell$
defined by the partition function $Z=\int\mathcal{D}[\phi] \exp (iS)$ with the action
\begin{align}
S=\frac{1}{4\pi\alpha'}\int dt \int_{0}^{\ell} dx
\left[
\frac{1}{v}( \partial_{t}\phi )^{2}
-v( \partial_{x}\phi )^{2}
\right],
\end{align}
where 
the spacetime coordinate of the edge theory is denoted by $(t,x)$,
$v$ is the velocity, 
$\alpha'$ is the coupling constant,
and  the $\phi$-field is compactified as
\begin{align}
\phi \sim \phi + 2\pi R,
\end{align}
with the compactification radius $R$. 
The canonical commutation relation is
\begin{align}
\left[
 \phi(x,t), \partial_t {\phi}(x^{\prime},t)
\right]
=
i
2\pi \alpha' v
\sum_{n\in \mathbb{Z}}\delta(x-x^{\prime}-n\ell).
\end{align}
We use the chiral decomposition of the boson field, and introduce the dual field $\theta$ as 
\begin{align}
\phi=\varphi_L+\varphi_R,
\quad
\theta =\varphi_L-\varphi_R. 
\end{align}
The mode expansion of the chiral boson fields is given by ($x^{\pm}=vt\pm x$) 
\begin{align}
\varphi_{L}(x^{+})
&=
x_L
+
\pi \alpha' p_L
\frac{ x^{+}}{\ell}
+
i 
\sqrt{ \frac{\alpha'}{2} }
\sum^{n\neq 0}_{n\in \mathbb{Z}}
 \frac{\alpha_{n}}{{n}}e^{-\frac{2\pi i nx^{+}}{\ell}}
, 
\nonumber \\
\varphi_{R}(x^{-})
&=
x_R
+
\pi \alpha'  p_R
\frac{x^{-}}{\ell}
 +
i \sqrt{ \frac{\alpha'}{2} }
\sum^{n\neq 0}_{n\in \mathbb{Z}}
 \frac{\tilde{\alpha}_{n}}{{n}}e^{-\frac{2\pi i nx^{-}}{\ell}} , 
\end{align}
where  
$ 
\left[\alpha_m, \alpha_{-n}\right] =  
\left[\tilde{\alpha}_m, \tilde{\alpha}_{-n}\right] =  
m\delta_{mn}$ and $ [x_L, p_L]= [x_R, p_R]= {i}$. 
The compactification condition on the boson fields 
implies that the allowed momentum eigenvalues are given by  
\begin{align}
&p=\frac{1}{2}\left(p_L + p_R\right) = \frac{k}{R},
\quad
\tilde{p} = \frac{1}{2} \left( p_L - p_R \right) = \frac{R}{\alpha'}w, 
\nonumber \\ 
&p_L = \frac{k}{R}+\frac{R}{\alpha'}w,  
\quad
p_R = \frac{k}{R}-\frac{R}{\alpha'}w,
\end{align}
where $k$ and $w$ are integers. 
In terms of these momentum eigenvalues, 
the compactification conditions on 
the boson fields are 
\begin{align}
&
\varphi_L(x+\ell) - \varphi_L(x)=+\pi \alpha' p_L,
\nonumber\\
&
\varphi_R(x+\ell) - \varphi_R(x)= -\pi \alpha' p_R,
\nonumber\\
&
\phi (x+\ell) - \phi(x) = \pi \alpha'(p_L-p_R)=2\pi R w,
\nonumber \\
&
\theta(x+\ell) - \theta(x) = \pi \alpha'(p_L+p_R) = 2\pi \frac{\alpha'}{R}k. 
\end{align}

The Hilbert space is constructed as a tensor product of 
the bosonic oscillator Fock spaces,  
each of which is generated by 
pairs of creation and annihilation operators 
$\{ \alpha_{m}, \alpha_{-m}\}_{m>0}$
and 
$\{\tilde{\alpha}_{m}, \tilde{\alpha}_{-m}\}_{m>0}$, 
and 
the zero mode sector associated with  ${x}_{L,R}$ and ${p}_{L,R}$.
We will denote states in the zero mode sector by specifying 
their momentum eigenvalues as
\begin{align}
 |p, \tilde{p}\rangle = | k/R, R w/\alpha'\rangle,
 \quad
 k, w\in \mathbb{Z},  
\end{align}
or more simply as $|k,w\rangle$. 
Alternatively, 
the Fourier transformation of the momentum eigenkets 
defines the ``position'' eigenkets, which we denote by
\begin{align}
 |\phi_0, \theta_0\rangle 
 \quad
 0 <\phi_0 \le 2\pi R,
 \quad
 0< \theta_0 \le 2 \pi \alpha'/R. 
\end{align}
The two bases are related by
\begin{align}
 | p, \tilde{p}\rangle
 =
 \int^{2\pi R}_0 d\phi_0
 \int^{2\pi \alpha'/R}_0 d\theta_0
e^{ -i p \phi_0 -i \tilde{p}\theta_0}
|\phi_0,\theta_0\rangle. 
\end{align}


The single-component compactified boson theory 
is invariant under various symmetry operations.
First of all, 
in the free boson theory, when there is no perturbation,   
there are two conserved $U(1)$ charges, one for each left- and right-moving sector.
Corresponding  to these conserved quantities, 
the free boson theory is invariant under the following 
$U(1)\times U(1)$ symmetry 
\begin{align}
{U}_{\delta\phi, \delta \theta}&:
 \phi \to \phi + \delta \phi,
 \quad
\theta \to 
\theta +
\delta \theta, 
\nonumber \\
&:\varphi_L 
\to
\varphi_L 
+
\delta \varphi_L, 
\quad 
\varphi_R 
\to
\varphi_R 
+
\delta \varphi_R, 
\end{align}
where $\delta \varphi_L = \frac{\delta \phi + \delta \theta}{2}$ and $\delta \varphi_R =\frac{\delta \phi - \delta \theta}{2}$. In terms of the conserved charges, 
the generators of the $U(1)\times U(1)$ transformations are given by 
\begin{align}
&
 \hat{U}^L_{\delta \varphi_L}
 =
e^{i \delta \varphi_L  
 N_{L}/(\alpha' \pi)
}
=
e^{ 
 i \delta \varphi_L 
 p_{L}
}, 
\nonumber \\
&
\hat{U}^R _{\delta \varphi_R}
 =
e^{ 
 i \delta \varphi_R
 N_{R}/(\alpha' \pi)
}
=
e^{ 
 i\delta \varphi_R p_{R}
},
\nonumber \\
&
\hat{U}_{\delta\phi,\delta\theta} =
\hat{U}^L_{\delta \varphi_L}
\hat{U}^R_{\delta \varphi_R}
=
e^{
 i (\delta \phi p + \delta \theta \tilde{p})}. 
\nonumber \\
& 
\mbox{where}\quad 
 N_{L,R}= \int^{\ell}_0  dx\, \partial_x \varphi_{L,R} =
\alpha' \pi p_{L,R}. 
\end{align}
%
Note that  
$\hat{U}_{\delta \phi, \delta \theta}$
acts on the momentum eigenkets as  
\begin{align}
\hat{U}_{\delta \phi, \delta \theta}  |p,\tilde{p}\rangle = 
e^{i (p\delta \phi + \tilde{p} \delta\theta)} |p, {\tilde p} \rangle. 
\end{align}

Another important symmetry in our discussion of the Haldane phase is 
particle-hole symmetry. 
Particle-hole symmetry or charge conjugation ($C$-symmetry) is unitary and acts on 
the bosonic fields as  
\begin{align}
{C}&:\phi \rightarrow - \phi + n_{c}\pi R,
\quad 
\theta \rightarrow -\theta + \frac{m_{c}\pi \alpha'}{R} \nonumber\\ 
&: (x_1, x_2) \rightarrow (x_1, x_2),
\label{csymmetry:phase}
\end{align}
where $(n_{c}, m_{c}) \in \{0,1\}$. 
From these transformation laws of the boson fields, 
we read off the action of $C$-symmetry on the position basis as
\begin{align}
 \hat{C}| \phi_0,\theta_0\rangle 
=
e^{i\delta}
 \left|-\phi_{0} + n_{c}\pi R, -\theta_{0} + m_{c}\pi \alpha'/R 
 \right\rangle,  
 \label{phase ambiguity, C-symmetry}
\end{align}
where $e^{i \delta}$ is an unknown phase factor. 
In order to have the relation 
$\hat{C}|p, \tilde{p}\rangle \propto |-p, -\tilde{p}\rangle$, 
expected from the commutation relation between $\hat{C}$ and $p,\tilde{p}$, 
the phase $\delta$ has to be a constant (independent of $\phi_{0}$ and $\theta_0$). 
The action of $C$-symmetry on the momentum eigenstates is given by 
\begin{align}
\hat{C} 
|p, {\tilde p} \rangle 
&= 
e^{i \delta}
e^{-ipn_{c}\pi R - i{\tilde p}\frac{m_{c}\pi \alpha'}{R}} 
|-p, -{\tilde p} \rangle 
\nonumber \\
&= 
e^{i \delta}
e^{-i\pi  kn_{c}  - i \pi w m_{c}} 
|-p, -{\tilde p} \rangle,  
\label{csymmetry:momentum}
\end{align}
where $p=k/R$ and $\tilde{p}=w R/\alpha'$. 
Since $\delta$ is constant, the phase ambiguity is fixed once we specify the action of 
$\hat{C}$ on a {\it reference state}, e.g., $|p,\tilde{p}\rangle = |0,0\rangle$.
In our analysis presented below, the reference state and its charge conjugation parity eigenvalue $e^{i\delta}$
plays an important role. 

%
%
%
%
%

Following Refs.\ \onlinecite{berg2010quantized, Fuji}
(see also Ref.\ \onlinecite{Bi2013}),
we now adopt the convention
\begin{align}
\alpha' = \frac{1}{2},
\quad
R = \frac{1}{2},
\quad
\frac{R}{\sqrt{\alpha'}} = \frac{1}{\sqrt{2}},
\quad
v=1.
\end{align}
This set of the parameters realizes the free fermion point in the 
moduli space of  $c=1$ CFTs.
(In our conventions the $SU(2)$ point (the self-dual radius) is realized 
when $R/\sqrt{\alpha'}=1$.)

In this convention, the Haldane phase
can be described by the 
sine-Gordon model: 
\begin{align}
H = \int dx 
\left\{
\frac{1}{2\pi}  \Big[ (\partial_x \theta)^{2} + (\partial_x \phi)^2  \Big] -\lambda \cos(2\phi)
\right\}, 
\label{BosonicHI}
\end{align}
where 
the bosonic fields are compactified as $\phi \sim \phi + \pi$ and $\theta \sim \theta + 2\pi$. 
The field $\theta$ is introduced to represent the canonical conjugate variable of the $\phi$, 
and 
the fields $(\phi, \theta)$ 
are related to (slow-modes of) the microscopic spin $(S_x (x), S_y (x), S_z(x))$ by 
\begin{align}
&S_x(x) + iS_y (x) \sim e^{i \theta (x)}, \nonumber\\ 
&S_z (x) \sim \partial_{x} \phi. 
\end{align}
Considering the Haldane model as a model of the (hard-core) boson, we can also relate the fields $(\phi, \theta)$ to (slow-modes of) the microscopic boson $b(x)$ and its density fluctuation $\delta \rho(x)$  
\begin{align}
&b(x) \sim \sqrt{\bar{\rho}} e^{i \theta (x)}, \nonumber\\ 
&\delta \rho(x) = \rho (x) -\bar{\rho} \sim \frac{1}{\pi}\partial_{x} \phi. 
\end{align}
where $\bar{\rho}$ is the average density of the boson. 

Under 
$\mathbb{Z}_2\times \mathbb{Z}_2$ symmetry
(the $\pi$ rotations of spins around $S_x$, $S_y$, and $S_z$-axis), 
the phase variables are transformed as
\begin{align}
R^x_{\pi}&: \phi\to -\phi, \quad \theta \to -\theta
\nonumber \\
R^y_{\pi}&: \phi\to -\phi, \quad \theta \to -\theta +\pi
\nonumber \\
R^z_{\pi}&: \phi\to \phi, \quad \theta \to \theta +\pi.
\end{align}
These transformation can be generated by combining 
the charge conjugation $\hat C$ and the $U(1)$ phase rotation $\hat{U}_{\delta\theta=\pi}$.
(On the other hand,
time-reversal acts on the phase fields as
$
T: \phi \to -\phi$,
$
\theta \to \theta + \pi
$.)

The cosine term $-\lambda \cos(2\phi)$ in Eq.\ \eqref{BosonicHI} 
is allowed by the symmetry 
(though it is not the only perturbation allowed by the symmetry). 
The theory \eqref{BosonicHI} 
describes the phase transition between the trivial Mott insulator and symmetry protected topological insulator, 
{\it i.e.,} the Haldane insulator.
\cite{berg2010quantized} 
The transition is triggered by changing sign of the coefficient $\lambda$ of the consine term in the effective theory Eq.\ \eqref{BosonicHI}. 

\subsection{The entanglement spectrum of the Haldane phase}


Following our general considerations,
we now discuss the BCFT description of the entanglement spectrum of the Haldane phase.
Setting $\lambda=0$ in Eq.\ \eqref{BosonicHI}
the relevant CFT is the single-component compactified free boson.  
To identity the relevant boundary conditions, 
let us first consider the two gapped phases
realized in the Hamiltonian \eqref{BosonicHI}, 
by taking $\lambda \to \infty$ and $\lambda \to -\infty$.
In the both phases, the cosine term strongly pins the $\phi$ field to its minima: 
For $\lambda \to \infty$, $\phi$ is pinned at $0$ mod $\pi$. 
On the other hand, for $\lambda \to -\infty$, $\phi$ is pinned at $\frac{\pi}{2}$ mod $\pi$. 

Let us next consider the domain wall between the two phases by changing $\lambda$ as a function of $x$. 
The domain wall is realized by the following configuration of $\lambda(x)$ :  
\begin{align}
\lambda(x) &=\left\{
\begin{array}{ll}
 - \Lambda & ~\text{for}~ x<0, 
\\ 
 0  & ~\text{for}~ x \in [0, \ell], 
\\ 
 + \Lambda & ~\text{for}~ x>\ell. 
\end{array}
\right.
\label{BC}
\end{align}
We will take the limit $\Lambda \to +\infty$ so that the theory of $x <0$ or $x>\ell$ 
is in its ground states of the cosine term of Eq.\ \eqref{BosonicHI}
with the corresponding sign of the coefficient $\Lambda$. 
(The conventional domain wall picture can be then realized by taking $\ell \to 0^{+}$.) 
Hence we effectively consider a critical  boson theory 
which is spatially sandwiched by the two topologically distinct insulator phases. 
We thus consider the boundary condition:
\begin{align}
\phi(x =0) &= \frac{\pi}{2}~ \text{mod}~\pi, \nonumber\\ 
\phi(x = \ell) &= 0 ~ \text{mod}~\pi. 
\end{align}


Before calculating the spectrum of the BCFT, and hence the entanglement spectrum, 
let us discuss the presence of the domain wall mode from somewhat complementary point of view.
We expect that there should be a zero mode, {\it i.e.,} a solitonic operator, 
in the critical regime, which is identified with the topological boundary modes of the Haldane chain. 
We would like to identify this soliton operator in the language of the CFT. 
For this, we need to look carefully into the boundary condition imposed on the boson field $\phi$. 
From the boundary condition, 
we find that 
\begin{align}
\int^{\ell}_{0} dx ~\frac{\partial_x \phi}{\pi}= \int^{\ell}_{0} dx ~\delta \rho(x) = n \pm \frac{1}{2}, \quad n \in {\mathbb Z}. 
\end{align}
Hence the soliton object we consider is created by
\beq
z_{\uparrow} \sim e^{i\frac{\theta}{2}}, ~\text{and}~ z_{\downarrow} \sim e^{-i \frac{\theta}{2}}. 
\label{BSpinor}
\eeq
Let us emphasize that the fields $z_{\sigma}, \sigma \in \{ \uparrow, \downarrow \}$ 
are the creation operators of the half charge of the fundamental boson and thus are the fractional degrees of freedom of the original boson. 
Furthermore, 
it is straightforward to check that 
they satisfy the projective symmetry group representation. 
Furthermore, it is now straightforward to check that the configuration  
\begin{align}
\lambda(x) &= 
\left\{
\begin{array}{ll}
+\Lambda & ~\text{for}~ x<0, 
\\ 
0 & ~\text{for}~ x \in [0, \ell],
\\ 
+ \Lambda &~\text{for}~ x>\ell, 
\end{array}
\right.
\label{BC_trivial}
\end{align}
has no non-trivial degenerate zero mode realizing the projective symmetry representation for both the limits $\Lambda \to \pm \infty$.

Depending on the sign of $\Lambda$, 
there are two drastically different behaviors of the spectrum in terms of topological degeneracy, which is  the focus of our interest. 
Hence we discuss the two cases separately. 
In general, the mode expansion of the boson field $\phi(t, x)$ is the following
\cite{CFTbook,PolchinskiStringbook}:
\begin{align}
\phi(t,x) = \frac{\Delta \phi - p \pi}{\ell} x + \sum_{n \in \mathbb{Z}, n \neq 0} \alpha_{n} \frac{e^{-2\pi i n t}}{n}  
\sin\left(\frac{2\pi n}{\ell}x\right),  
\label{mode}
\end{align}
in which $p \in \mathbb{Z}$ determines the winding of the bosons, 
$\Delta \phi = \phi(x = \ell)-\phi(x=0)$ mod $\pi$ to be determined by the boundary conditions, 
and $\alpha_n$ is the harmonic oscillator satisfying 
$[\alpha_n, \alpha_m] = m \delta_{n+m,0}$. 
The entanglement Hamiltonian in terms of the mode decomposition can be written as:
\cite{CFTbook,PolchinskiStringbook}
\begin{align}
H = \frac{2}{\ell}\left[
\left(\frac{\Delta \phi}{\pi} - p\right)^2 + \sum_{n>0} \alpha_{-n}\alpha_n 
\right], ~ p \in \mathbb{Z}.
\end{align}
When $\Delta \phi=0$, 
the lowest state of the tower for $p =0$ 
is non-degenerate and so are all states in the tower of states.
Thus the entanglement spectrum is trivial. 
On the other hand, when $\Delta\phi=\pi/2$, 
the lowest states of the tower $p=0$ and $p=1$ are degenerate,
and all states in the spectrum are at least doubly-degenerate. 
Furthermore, 
by state-operator correspondence, 
the two lowest states corresponds to the spinor \eqref{BSpinor},
which transform projectively under symmetry. 
Thus the degeneracy in the entanglement spectrum is protected 
by symmetry as exactly the same way as the physical boundary zero modes. 

\subsection{Boundary states}
\label{Boundary states, Haldane phase}

Let us now use the boundary states
to show (again) the symmetry-protected degeneracy.
We will also derive the anomalous phase of the boundary 
state in the twisted sector.

The boundary state with $\phi(0)=\phi_0$ can be explicitly constructed as 
\begin{align}
|D(\phi_0)\rangle
&=
\sqrt{ \frac{1}{R\sqrt{2} } }
\exp\left( \sum_{n=1}^{\infty}
\frac{1}{n}\alpha_{-n} \tilde{\alpha}_{-n} 
\right)
\nonumber \\
&\quad
\times 
\sum_{p=k/R, k \in \mathbb{Z} }
e^{ -i p  \phi_0 } |p, 0\rangle. 
\end{align}
This state is invariant under 
$
{U}_{ \delta \theta} 
$
and
$
{C}. 
$
The partition function can be computed from the boundary state as
\begin{align}
\langle D(\phi_0) |
\tilde{q}^{\frac{1}{2} (H_L+H_R) }
|D(\phi'_0)\rangle
=
\frac{1}{\eta(\tilde{q})} 
\sum_{m\in \mathbb{Z}} q^{ (m+1/2)^2}.
\end{align}
This spectrum shows that 
all states are at least doubly degenerate. 

\paragraph{$R^z_{\pi}$-twisted sector}

Following our general discussion, we now consider boundary states in twisted sectors.
In particular, we will confirm the symmetry-enforced vanishing of the partition function,  
by computing the anomalous phase of boundary states that 
may be picked up under the action of symmetry.
Let us first now consider 
the twist by $R^z_{\pi}$
\begin{align}
\theta(x+\ell) = \theta(x) + 2\pi \alpha' k/R + \pi \alpha' /R, 
\end{align}
where $k$ is an integer.
With this twist, the allowed momentum is now
\begin{align}
p = \frac{1}{2} (p_L + p_R) = \frac{1}{R} (k+1/2), 
\end{align}
as one can see from the mode expansion of the boson fields. 
The boundary state with $\phi(0)=\phi_0$ in the presence of the twist
is 
\begin{align}
|D(\phi_0)\rangle_{R^z_{\pi}}
&=
\sqrt{ \frac{1}{R\sqrt{2} } }
\exp\left( \sum_{n=1}^{\infty}
\frac{1}{n}\alpha_{-n} \tilde{\alpha}_{-n} 
\right)
\nonumber \\
&\quad
\times 
\sum_{p=(k+1/2)/R, k \in \mathbb{Z} } e^{ -i p \phi_0 /R } |p, 0\rangle. 
\end{align}
When $\phi_0= \pi R$, 
the symmetry $\hat{C}$ acts on the boundary state as 
\begin{align}
&
\hat{C} \sum_{p=\frac{1}{R}\left(k+\frac{1}{2}\right), k \in \mathbb{Z} } e^{ -i p \phi_0 /R } |p, 0\rangle. 
\nonumber \\
&= 
\sum_{p=\frac{1}{R}\left(k+\frac{1}{2}\right), k \in \mathbb{Z} } 
e^{ -i p \phi_0 /R } |-p, 0\rangle. 
\nonumber \\
&=
\sum_{p=\frac{1}{R}\left(k+\frac{1}{2}\right), k \in \mathbb{Z} } 
e^{ +i p \phi_0 /R } |p, 0\rangle. 
\end{align}
Since
$
e^{ i p \phi_0/R}= 
e^{i (k+1/2) \pi} =
(-1)e^{-i (k+1/2) \pi}
=
(-1)e^{-i p \phi_0/R}, 
$
we conclude that the 
boundary state picks up a minus sign under the action of 
$\hat{C}$: 
\begin{align}
\hat{C}|D(\pi R)\rangle_{R^z_{\pi}}
=
- |D(\pi R)\rangle_{R^z_{\pi}}. 
\end{align}
We thus conclude the corresponding partition function
is forced to zero due to symmetry.

\paragraph{$R^x_{\pi}$-twisted sector}

Let us now consider orbifolding by $R^x_{\pi}$:
\begin{align}
\phi(x+\ell) &= -\phi(x) + 2\pi R n,
\nonumber \\
\theta(x+\ell) &= -\theta(x) + 2\pi \frac{\alpha'}{R} m,
\end{align}
where $n$, $m$ are some integers. 
This twist sets the momentum to be zero, $p=\tilde{p}=0$, and 
the mode expansion compatible with the twist is given by  
\begin{align}
\phi(x) = x_L + x_R + \cdots 
\end{align}
where $\cdots$ represents oscillator modes. 
In the twisted sector the zero mode $x_L+x_R$ can only take its fixed point value
$0$ or $\pi R$. 
Thus, there are two independent states in the zero mode sector,
$|0\rangle_{R^x_{\pi}}$ and $| \pi R \rangle_{R^x_{\pi}}$.
(See, for example, Refs.\ \onlinecite{Oshikawa1997, Cappelli2013}.)
The Dirichlet boundary states can then be constructed from the states as
\begin{align}
|B(\phi_0) \rangle_{R^x_{\pi}} \propto \exp (\mbox{oscillator part})| \phi _0 \rangle_{R^x_{\pi}},
\end{align}
where $\phi_0 = 0$ or $\phi_0 = \pi R$. 

In addition to the Dirichlet boundary states in the twisted sector,
the orbifold theory twisted by $R^x_{\pi}$ allows the boundary states in 
the untwisted sector. They are simply given by a suitable linear combination of the 
boundary states which are invariant under $R^x_{\pi}$. 
These boundary states represents boundary conditions in which the boson field 
is pinned at a certain value $\phi_0$, where $\phi_0$ can be arbitrary. 
When $\phi_0$ is at the fixed points of the symmetry, $\phi_0=0$ or $\phi_0=\pi R$, 
these untwisted boundary states by themselves fail to satisfy the Cardy condition. 
It is then necessary to consider the boundary states in the twisted sector considered above. 

The two zero mode states $|\phi_0\rangle_{R^x_{\pi}}$, 
and hence the two boundary states $|B(\phi_0)\rangle_{R^x_{\pi}}$, 
are orthogonal to each other, and hence the partition function 
\begin{align}
{ }_{R^x_{\pi}}\langle B(0) |
\tilde{q}^{\frac{\ell}{2} H^{closed} }
|B(\pi R)\rangle_{R^x_{\pi} }
=0
\end{align}
vanishes. 
To see if this is symmetry enforced, 
we need to consider the action of symmetry on these boundary states, 
say, $R^z_{\pi}$.
I.e., 
$R^z_{\pi} |B(\phi_0)\rangle_{R^x_{\pi}}$.
To this end, let us first consider Neumann boundary states in the twisted sector.
They are given by
\begin{align}
| N(\theta_0)\rangle_{R^x_{\pi}} = e^{(\mbox{osc. part})} \frac{1}{\sqrt{2} } \left(
|0\rangle_{R^x_{\pi} } \pm | \pi R\rangle_{R^x_{\pi}} 
\right). 
\end{align}
where $\theta_0=0$ or $\theta_0 = \pi \alpha'/R$. 
Since $R^z_{\pi}$ shifts $\theta$ by $\pi$, we expect that $R^z_{\pi}$ exchanges $|N(0)\rangle_{R^x_{\pi}}$ 
and $|N(\pi \alpha'/R) \rangle_{R^x_{\pi}}$. 
That is,
\begin{align}
R^{z}_{\pi} | D(0) \rangle_{R^x_{\pi} } &= |D(0) \rangle_{R^x_{\pi}}, 
\nonumber \\
R^{z}_{\pi} | D(\pi \alpha'/R) \rangle_{R^x_{\pi} } &= -|D(\pi \alpha'/R) \rangle_{R^x_{\pi}}, 
\end{align}
up to a possible common over all phase. 
The anomalous minus sign picked up by $|D(\pi\alpha'/R)\rangle_{R^x_{\pi} }$ under $R^z_{\pi}$
shows the vanishing of the partition function is enforced by the  symmetry of the Haldane phase. 
%
%
%
%
%
%
%
%
%
%

\section{(1+1) d topological superconductors in  symmetry class BDI}
\label{BDI topological superconductors in (1+1)d}

In this section, 
we consider topological superconductors in symmetry class BDI,
and the $\mathbb{Z}_8$ classification of Fidkowski and Kitaev\cite{Fidkowski2010,Fidkowski2011}.
Following our general framework, we will use BCFT to detect the $\mathbb{Z}_8$ classification. 
Our analysis in terms of boundary states 
gives an alternative perspective of the $\mathbb{Z}_8$ classification of Fidkowski and Kitaev
 in terms of quantum anomalies of boundary states of CFT.

We emphasize that, 
in our analysis below, 
we will use boundary states in {\it free} fermion CFTs
to detect the $\mathbb{Z}_8$ classification in class BDI,
which
arises from the
reduction
 of the $\mathbb{Z}$ classification
in the presence of interactions. 
While all calculations will be done here entirely within the context of free fermion manipulations,
nevertheless, 
it should be noted that
(i) boundary states 
are constructed in the many-body Hilbert space (the Fock space).
Moreover, 
(ii)    
anomalous phases that boundary states may acquire upon 
the action of symmetry operations 
are  expected to ''survive``
or to ''be protected``, even in the presence of interactions
(see Sec.\ \ref{Boundary conditions and anomalous phases}
for related discussion), 
in analogy to 
various kinds of quantum anomalies in quantum field 
theories.

We also note that technically,
the following discussion has much resemblance to  
the analysis of quantum anomalies at the edge of
(2+1)-dimensional topological crystalline superconductors,
for which the classification is $\mathbb{Z}_8$.
\cite{yao2013interaction}
In Ref.\ \onlinecite{ChoHsiehMorimotoRyu2015},
a quantum anomaly of the corresponding (1+1)-d edge theory was identified 
to diagnose the $\mathbb{Z}_8$ classification
by using cross-cap states in CFTs.
The boundary states discussed in this section of the present paper, 
when restricted to the zero-mode sector of the closed-string Hilbert space,
are 
identical to those appearing in  the  cross-cap states that arose in the analysis of the (1+1)-d edges, which
are obtained by ``gauging" 
a  mirror (or ``reflection'',  or ``parity'')  symmetry. In fact, this is consistent with the fact that the classification of non-interacting
 (2+1)-dimensional topological insulators and topological superconductors with mirror symmetry is identical to that of 
non-interacting (1+1)-dimensional topological insulators and superconductors without mirror symmetry\cite{Chiu2013}. Our analysis presented in the present work, based on a quantum anomaly of boundaries of (1+1)-d gapped SPT phases, 
implies that the classification for (2+1)-dimensional topological insulators and superconductors with mirror symmetry is the same as that of (1+1)-dimensional topological insulators and superconductors without mirror symmetry even in the presence of the interactions, where 
a certain $\mathbb{Z}$ classification, such as that occurring in symmetry class BDI, is reduced to a $\mathbb{Z}_8$ classification. In this section, we will particularly be interested in the 
 BDI class.

Consider the  CFT  consisting of $N_f$ flavors of non-interacting  non-chiral (i.e. right and left moving)  real (Majorana) fermions
described by the Hamiltonian \eqref{fermionic edge theory}.
In addition to fermion number parity conservation, 
we impose
on the system
 time-reversal symmetry 
\begin{align}
\label{LabelEqTimeReversalBDI}
&
\hat{T} \psi^a_L(t,x) \hat{T}^{-1} = \psi^a_R(-t,x),
\nonumber \\
&
\hat{T} \psi^a_R(t,x) \hat{T}^{-1} = \sigma \psi^a_L(-t,x), 
\nonumber \\
& \quad \hat{T}^2=\sigma^F,
\  \ 
\hat{T}i \hat{T}^{-1} = -i,
\end{align}
where $\sigma=\pm 1$,
and  $a=1, ..., N_f$. 
The fermion parity operator $F$ was defined in Eq.\ (\ref{LabelDEFTotalFermionParitityNumberOperator}).
 The case of interest
for us is $\sigma=+1$, relevant for symmetry class BDI.
%

To discuss the action of time-reversal on boundary states of the corresponding free fermion CFT, 
we will now implement the  $\pi/2$ rotation of Euclidean (i.e., imaginary time)  spacetime,
discussed already in the paragraphs surrounding  Eqs.\ (\ref{LabelEqBoundaryConditionsOnFields})
and (\ref{LabelEqClassDPiOverTwoRotation}). Since this rotation involves imaginary (Euclidean) time, we first
need to reformulate the condition of time reversal invariance, Eq.\ (\ref{LabelEqTimeReversalBDI}), as a condition
involving imaginary time. This can be understood, e.g.,  by  using the fact that the non-interacting Majorana fermion theory in
  \eqref{fermionic edge theory}
satisfies the CPT Theorem (since this is  a theory of   Lorentz invariant Dirac/Majorana fermions).

In the present case of Majorana fermions, charge-conjugation $C$ acts trivially, and therefore the condition of time-reversal 
invariance in Eq.\ (\ref{LabelEqTimeReversalBDI}) is satisfied if and only if the following condition of ``parity'',
 or equivalently ``spatial reflection'' ${\cal R}$
symmetry is satisfied
\begin{align}
\label{LabelEqParity}
&
\hat{\cal R} \psi^a_L(t,x) \hat{\cal R}^{-1} = \psi^a_R(t,\ell - x),
\nonumber \\
&
\hat{\cal R} \psi^a_R(t,x) \hat{\cal R}^{-1} = (- \sigma) \ \psi^a_L(t,\ell-x), 
\nonumber \\
& \quad \hat{\cal R}^2=(-\sigma)^F,
\  \ 
\hat{\cal R}i \hat{\cal R}^{-1} = i \  ({\rm unitary}),
\end{align}
where we considered the situation where the fermions are defined on a spatial  circle of circumference $\ell$
 with coordinate $x$.  It can be verified\footnote{See e.g. Ref. [\onlinecite{ChoHsiehMorimotoRyu2015}].}
 (e.g. by checking that this forbids the same mass terms) that one needs to
change the sign  $\sigma \to (-\sigma)$ as indicated,
when going from Eq.\ (\ref{LabelEqTimeReversalBDI}) to Eq.\ (\ref{LabelEqParity}).

Note that since ${\hat {\cal R}}$  in  Eq.\ (\ref{LabelEqParity}) acts only the spatial coordinate $x$, the same equation
holds true when  real time $t$ is replace by imaginary (Euclidean) time $\tau$, i.e. $t \to$ $\tau$, in that equation.
The imaginary (Euclidean) time version of time reversal from Eq.\ (\ref{LabelEqTimeReversalBDI}) is then the same equation as Eq.\ (\ref{LabelEqParity}), after the  rotation by $\pi/2$ of  (Euclidean) spacetime,  $(x,\tau)=$ $  (-{\tilde \tau}, {\tilde x})$,
which was already
discussed  in the paragraph 
surrounding Eq.\ (\ref{LabelEqBoundaryConditionsOnFields}),
 is implemented.
Denoting the Euclidean-time version of time-reversal by ${\hat P}$ (standing for ``parity''), the imaginary time version of
time reversal symmetry  reads
\begin{align}
\label{LabelEqEuclideanTimeReversal}
\hat{P} \psi^a_L({\tilde \tau}, {\tilde x}) \hat{P}^{-1} &=  \psi^a_R({\tilde \tau},\beta -  {\tilde x}),
\nonumber \\
 \hat{P} \psi^a_R({\tilde \tau}, {\tilde x}) \hat{P}^{-1} &= (- \sigma) \ \psi^a_L({\tilde \tau}, \beta - {\tilde x}), 
\nonumber \\
 \quad \hat{P}^2=(-\sigma)^F,&
\  \ 
\hat{P}i \hat{P}^{-1} = i \  ({\rm unitary}).
\end{align}
As a brief check, note that the simple free fermion boundary conditions in Eq.\ (\ref{LabelEqFreeFermionBoundaryConditions})
characterized by numbers $\eta=\pm 1$ are invariant under the time reversal transformation defined in  Eq.\ (\ref{LabelEqTimeReversalBDI})
when $\sigma=+1$ (the sign relevant for class BDI) as expected. Equivalently, the boundary state $|B(\eta)\rangle$ defined in Eq.\ (\ref{LabelEqDEFBoundaryStates}), can be written in the form  of Eq.\ (\ref{LabelEqBoundaryConditionsOnFields})
with
\begin{align}
\label{LabelEqBoundaryStateBDIVectorForm}
{\hat {\bf \Phi}}({\tilde x}) \equiv 
\left ( 
\begin{matrix}
\psi_L({\tilde x}) \\
\psi_R({\tilde x})
\end{matrix}
\right )
\quad {\rm and} \quad
 U =
i \eta 
\left ( 
\begin{matrix}
0 & -1 \\
+1 & 0 
\end{matrix}
\right ).
\end{align}
Following the steps in Eq.\ (\ref{LabelEqGroupInvarianceOfBoundaryCondition}) we find that this  boundary condition
preserves the symmetry ${\hat P}$ (i.e. time reversal), except that we still need to discuss the action of ${\hat P}$ on the
boundary state itself,  i.e. ${\hat P} |B\rangle_h$. This will be done in detail below.

We will now implement the analysis of Sec.\ \ref{LabelSection-SymmetryProtectedDegeneracyEntanglementSpectrumGeneralConsiderations},
specifically Sec.\ \ref{LabelSectionSymmetryEnforcedVanishing}. The discrete symmetry group of the current problem is
$\mathbb{Z}_2^F\times\mathbb{Z}_2^T$ generated by fermion parity ${\hat g}_f$ (as in Sec.\ \ref{The Kitaev chain}) and
time reversal symmetry, for which we use the formulation in terms of ${\hat P}$, as in  Eq.\ (\ref{LabelEqEuclideanTimeReversal}).
These two symmetry operations commute. Following Sec.\ \ref{LabelSectionSymmetryEnforcedVanishing},
we choose to twist the boundary state (i.e. we twist in  imaginary time $\tau={\tilde x}$) by the fermion parity operator ${\hat g}_f$
in the same way as was  done in  Sec.\ \ref{The Kitaev chain}.

It follows from
Eq.\ (\ref{LabelEqEuclideanTimeReversal}) that the fermion zero modes, Eq.\ (\ref{LabelEqDEFTwistedBoundaryStateMomentum}),
satisfy 
\begin{align} \nonumber
\hat{P} \psi^a_{sL} \hat{P}^{-1} &= e^{i 2\pi s} \psi^a_{sR},
 \\ \label{LabelEqParityFourierModes}
 \hat{P} \psi^a_{sR} \hat{P}^{-1} &=  (-\sigma) e^{-i 2\pi s}  \psi^a_{0L},
\\ \nonumber
(s\in \mathbb{Z}+{1\over 2},&  \ \  {\rm or} \ \ s\in \mathbb{Z}). 
\end{align}
This implies that ${\hat P}$, when acting on the boundary state twisted by fermion number parity,  $|B(\eta)\rangle_{{\hat g}_f}$ in Eq.\ (\ref{LabelEqExplicitFormBoundaryState}), commutes with the exponential in that equation when $\sigma=+1$ (relevant for class BDI);
therefore, 
we only need to discuss the action of ${\hat P}$ on the ``zero-mode contribution'' $|B (\eta)\rangle_0$ defined in
Eq. (\ref{LabelEqExplicitFormBoundaryState}). Let us denote by ${\hat P}_0$ the projection of the operator ${\hat P}$
on the zero-mode sector.
It can be verified by direct calculation  that 
an explicit expression 
satisfying
\begin{align}
\label{LabelEqParityZeroModes}
\hat{P}_0 \psi^a_{0L} \hat{P}_0^{-1} = \psi^a_{0R},
 \quad
 \hat{P}_0 \psi^a_{0R} \hat{P}_0^{-1} =  -\psi^a_{0L}, 
\end{align}
as required by Eq. (\ref{LabelEqParityFourierModes})
is given by
\begin{align}
\hat{P}_0 =
 e^{i\delta}
 \prod_{a=1}^{N_f} 
 \frac{1}{\sqrt{2}}
 \left(
 1-
 \psi^a_{0L}\psi^a_{0R}
 \right),  
\end{align}
where $e^{i\delta}$ is a so-far unknown phase factor which will be discussed in more detail 
shortly.
Moreover, one also verifies that
\begin{align}
\hat{P}^2 = e^{2i\delta} (i)^{N_f}
\hat{g}_f,
\quad
\mbox{and}
\quad 
 \hat{g}_f \hat{P}= \hat{P}\hat{g}_f
\label{projective}
\end{align}
on the zero mode sector.

Let us now calculate the action of $\hat{P}_0$ on
the zero-mode contribution $|B(\eta)\rangle_0$ of 
the boundary state (Eq. (\ref{LabelEqExplicitFormBoundaryState})).
By using the representation in terms of the $f$-fermions, 
$\psi_{0L}\psi_{0R} = i (2 f^{\dag} f - 1)$, 
$\hat{P}_0$ can be written as
\begin{align}
\hat{P}_0 =
e^{i\delta} 
 \prod_a 
 \frac{1}{\sqrt{2}}
 \left[
 1- i (2 n_a -1)
\right],  
\end{align}
where $n_a = f^{\dag}_a f_a$. 
Then, the action of $\hat{P}_0$ on the zero-mode part of the boundary states
is given by 
\begin{align}
\hat{P}_0 |B, +\rangle_0
&=
e^{i\delta}
 \prod_a 
 \frac{1}{\sqrt{2}}
 \left[
 1- i (2 n_a -1)
\right]  
|0_f\rangle
\nonumber \\
&=
e^{i\delta}
 \prod_a 
 \frac{1}{\sqrt{2}}
 \left[
 1+ i 
\right]  
|0_f\rangle
=
e^{ +i\frac{\pi}{4}N_f}
e^{i\delta}
|B, +\rangle_0, 
\nonumber \\
\hat{P}_0 |B, -\rangle_0
&=
e^{i\delta}
 \prod_a 
 \frac{1}{\sqrt{2}}
 \left[
 1- i 
\right]  
|B, -\rangle_0
=
e^{ -i\frac{\pi}{4}N_f} 
e^{i\delta}
|B, -\rangle_0. 
\end{align}
The relative phase between 
$\hat{P}_0|B, +\rangle_0$
and 
$\hat{P}_0|B, -\rangle_0$
is $e^{ + i \pi N_f/2}$, 
which 
is independent of the choice of $e^{i\delta}$ (the choice of the action of $\hat{P}_0$ on the reference state),
and vanishes  when $N_f = 4 \times \mbox{integer}$. 
In other words, 
one cannot make both boundary states anomaly-free unless $N_f = 4\times \mbox{integer}$. 
One then immediately concludes the classification is {\it at least} $\mathbb{Z}_4$.

On the other hand,
a proper choice of the phase $e^{i\delta}$, if it exists, leads to a refined classification as we will now demonstrate. 
If we choose $|0_f\rangle$ as the reference state and demand  that $|0_f\rangle$ transform trivially under $\hat{P}_0$, 
i.e. 
$\hat{P}_0|0_f\rangle = |0_f\rangle$, 
we obtain  a $\mathbb{Z}_4$ classification. 
To discuss a proper phase choice, we consider the following alternative construction 
of the boundary state.
When $N_f=\mbox{even}$, 
one can introduce the following fermion creation operators 
(see, for example, Ref. \onlinecite{Bergman1997}):
\begin{align}
d^{\dag}_{L j} = 
\frac{1}{{2}}(\psi^{2j-1}_{0L} + {i} \psi^{2j}_{0L}), 
\quad
{d}^{\dag}_{R j} =
\frac{1}{{2}}({\psi}^{2j-1}_{0R} + {i} {\psi}^{2j}_{0R}), 
\end{align}
and the Fock vacuum $|0_d\rangle$ annihilated by both, $d_{Lj}$ and $d_{Rj}$.  
We observe that
$
d_{Lj}- {i}{d}_{Rj}
=
f_{2j-1} - {i} f_{2j} 
$,
$(d^{\dag}_{jL} - i d^{\dag}_{jR}) |0_f\rangle=0$, 
and
\begin{align}
 |B, + \rangle_0 = |0_f\rangle = \prod_j (d^{\dag}_{jL} - i d^{\dag}_{jR}) |0_d\rangle. 
\end{align}
Similarly,
\begin{align}
|B, - \rangle_0 = \prod_{a=1}^{N_f} f^{\dag}_a  |0_f\rangle = \prod_j (d^{\dag}_{j R} - i d^{\dag}_{j L}) |0_d\rangle.
\label{2} 
\end{align} 
(A similar construction is possible also  for $N_f=\mbox{odd}$
by adding an extra Majorana fermion as in Ref. \onlinecite{Fidkowski2011}.
We already know from the discussion in Sec. \ref{The Kitaev chain}, by using only fermion parity and not
time reversal symmetry, that all cases $N_f=\mbox{odd}$ are topologically non-trivial. On the other hand,
the anomalous phase can be computed for $N_f=\mbox{odd}$ in a manner similar to $N_f=\mbox{even}$, but
we do not present explicit results here.)
One important feature of this construction is the clear factorization of
the vacuum $|0_d\rangle$ into the left- and right-moving sectors (of zero modes), 
$|0_d\rangle = |0_d\rangle_L\otimes |0_d\rangle_R$,
as it is annihilated by $d_{Lj}$ and $d_{Rj}$ separately.
The entire Hilbert space built out of $|0_d\rangle$ also factorizes into the left- and right-moving sectors.
This factorization allows us to introduce the action of 
 parity (time-reversal) in a transparent way. 
Such a factorization is also expected 
from the  general construction of 
boundary states 
 in boundary conformal field theories. 
(See, for example, Ref. \onlinecite{Bates2006}.)

If we choose $|0_d\rangle$ as the reference state, 
and fix the phase ambiguity by demanding $\hat{P}_0|0_d\rangle=|0_d\rangle$,
which may be obtained from
$\hat{P}_0|0_d\rangle_{L,R} = |0_d\rangle_{R,L}$. Then, 
\begin{align}
 \hat{P}_0|B, + \rangle
 &= \hat{P}_0 |0_f\rangle \nonumber \\
 &= \hat{P}_0 \prod_{j=1}^{N_f/2} (d^{\dag}_{jL} - i d^{\dag}_{jR} )|0_d\rangle
 \nonumber \\
 &=
 \prod_{j=1}^{N_f/2} (d^{\dag}_{jR} +i d^{\dag}_{jL} )|0_d\rangle
\nonumber \\ 
 &=
 (i)^{N_f/2}|0_f\rangle. 
\end{align}
I.e., $e^{i\delta} = 1$.
It can also be checked, straightforwardly,  
\beq
\hat{P}_0|B, - \rangle_0 =  (i)^{N_f/2} |B, - \rangle_0, 
\eeq  
following Eq.\ \eqref{2}. 
Thus, with this choice, the two boundary states $|B,\eta\rangle$ can be both made anomaly free
only when $N_f=8\times \mbox{integer}$,
which tells us that there is a  $\mathbb{Z}_8$ classification. 

As a final comment, we provide yet another point of view by using Eq.\ (\ref{projective}). 
Equation (\ref{projective}) suggests that, within the zero mode sector, the symmetry is realized projectively. 
The ``unwanted'' phase $e^{2i \delta} (i)^{N_f}$ can be removed by  
choosing $e^{i\delta} = e^{- i \pi N_f/4}$.
However, with this choice, the reference state now acquires an anomalous phase 
$\hat{P}_0|0_d\rangle = e^{ -i \pi N_f/4} |0_d\rangle$. 
This conflict between the two demands,
one to represent the symmetry group non-projectively 
and 
the other to make the reference state transform trivially under $\hat{P}$, 
can be considered as a form of quantum anomaly.

\section{Discussion}
\label{Discussion}

%
%
%
%

%


In this paper,  we have given a  description
of the entanglement spectrum of SPT phases in (1+1) dimensions which are in
vicinity of a quantum critical point described by a CFT, in terms of
a boundary CFT  (BCFT) associated with that describing the quantum critical point.
We also introduced a diagnostic tool, the symmetry-enforced vanishing of the twisted partition function,
which allows us to 
identify the presence of a non-trivial cocycle, and of  a projective representation 
of the symmetry group
defining the SPT phase in the entanglement spectrum,
and to identify the topological class of the SPT phases.
From the perspective of CFTs, our formalism allows us to identify SPT phases that can be proximate to 
a given  CFT.
Hence, it gives us the structure of the phase diagram (the `theory space') around the CFT. 
As yet another perspective, our formalism can be thought of as a proper generalization of the Jackiw-Rebbi soliton
from  non-interacting fermion systems to generic SPT phases. 

While we have made a connection between boundary states in CFTs on one hand,  and SPT phases
on the other,  it should be emphasized, again, that the correspondence is not one-to-one: 
Many different BCFTs can  correspond to a given SPT phase. 
For example, for a Haldane system,  one can attach  as an ideal lead
the $c=1$ compactified boson, the
$SU(2)_2$,
or the
$SU(3)_1$ CFTs. All these CFTs are proximate to the Haldane phase, in the sense discussed in this paper, i.e., 
 they describe, respectively,  three possible  (conformal) quantum critical points through which one
can exit the Haldane phase into other (typically non-topological) phases.
Therefore, while one should be able to use boundary states of these CFTs to diagnose the
topological properties of the Haldane phase,
when it comes to classifying SPT phases, it is not optimal to use BCFTs, because
several different BCFTs (the three mentioned above and also others) can be
used to describe the same SPT phase (here the Haldane phase).
In this sense, our BCFT approach is complementary to other approaches, such as e.g. the MPS approach. 
In fact, the complete classification of boundary states in CFTs so far has not been achieved, 
while (1+1)d SPT phases are completely classified by $H^2(G, U(1))$. 
In other words, boundary states in CFTs or BCFTs seem to have ''too much information``;
(I.e., the set of all boundary states seem much bigger than  the set of all possible SPT phases in (1+1)d.). 

Another issue which may be related to this is the difference between symmetry-protected degeneracy 
in the entanglement spectrum and  the boundary entropy.
The former degeneracy should exist at all length scales,
while the latter should emerge only in the long-wave length limit,
and looks like a much more non-trivial property than the symmetry-protected degeneracy. 
Since when describing (1+1)-dimensional SPTs we are  interested in the properties of BCFTs which are
in fact  independent of the length scale, 
there may be an efficient way to extract this topological information out of the  BCFTs.   
One can speculate that such information can be extracted by a procedure
such as the
''topological twist``, when applicable. 
Such a procedure essentially turns (B)CFTs into topological field theories, and hence
the resulting (``topologically twisted'')  theory will only contain
information that is independent of the length scale.
In other words, such a  hypothetical procedure should ''remove`` the  unnecessary information from the CFT,
so that only topologically relevant information remains. 
Along this line of thought, we note that the complete classification of boundary states in (1+1)d
topological quantum field theories (TQFTs)  is  actually much more well understood than in  BCFTs, and has been studied, e.g.,
in the work of Moore and Segal.
In short, we conjecture that there is a close connection between SPT phases and boundary states in 
(1+1)d TQFTs. 

Finally, we end by mentioning that the appearance of a defect CFT which appears as an interface between
two different (1+1) d CFT connected via a gapped SPT region provides an interesting generalization of the set up
discussed in this paper. However, we leave this topics for future work.

\acknowledgements

We thank 
Anton Akhmerov, 
John Cardy,
Kiyonori Gomi,  
Kentaro Hori, 
Chang-Yu Hou,
Kantaro Ohmori,
Masaki Oshikawa, 
Xiaoliang Qi, 
and
Mike Zaletel,
for helpful discussion. 
We are grateful to the 
KITP Program ``Entanglement in Strongly-Correlated Quantum Matter''
(Apr 6 - Jul 2, 2015),
and 
YITP Long term workshop 2016
``Quantum Information in String Theory and Many-body Systems''
(May 23 - June 24, 2016).
This work is supported by the NSF under Grants 
No.\ NSF PHY11-25915,
No.\ DMR-1064319 (G.Y.C.) 
No.\ DMR-1455296 (S.R.),
and 
No.\ DMR-1309667 (A.W.W.L.),
the Brain Korea 21 PLUS Project of Korea Government (G.Y.C),
JSPS Postdoctoral Fellowships for Research Abroad (K.S.), 
and 
Alfred P. Sloan foundation (S.R.). 

\vspace{.3cm}

{\it Note added}: After  the key  results
of the work reported here were obtained, 
a preprint, Ref.\ \onlinecite{Tsui2015}, which discusses 
the properties of CFTs that may appear 
as a continuous quantum phase transition between two
different SPT phases sharing the same symmetry. 
Our analysis concerning the entanglement spectrum of SPTs,
which is described by BCFTs (as opposed to bulk CFTs),
shares some similarities with Ref.\ \onlinecite{Tsui2015}.

\appendix

\section{Fractional branes and discrete torsion}

In this Appendix we would like to
make a few historical remarks.
In the context of D-branes in string theory, 
it was observed that open string states in the 
open string channel
(i.e., states in BCFTs in the 
open string channel)
may form a projective representation of an orbifold group $G$, as in Eq.\ \eqref{proj. rep}. 
\cite{Douglas1998, Douglas1999, Sharpe2003, Billo2001}
(Typically, this statement is phrased in terms of spacetime quantum fields living on D-branes.)

Boundary states that are relevant to SPT phases are
those that are invariant (up to an anomalous phase) under the action of the symmetry 
defining the SPTs.
In the terminology of orbifold CFTs and D-branes, 
they are D-branes that are localized at orbifold fixed points.
Such D-branes are called fractional branes.
Fractional branes which may exist in a theory with discrete torsion have been also discussed.  
(Here, the theory means string theory which includes both open and closed string, and describes interactions among them.) 

In the context of fractional branes, 
it was argued that
a  two-cocyle that appears in the action of orbifold group $G$ in the 
open string channel
is directly related to the discrete torsion that appears in the closed string Hilbert space.
As the notation suggests, 
if $\omega(g|h)$ is the two-cocycle that appear in the open string picture, 
it was argued that 
$\varepsilon(g|h)$ in Eq.\ \eqref{discrete torsion and two-cocyles} 
is nothing but the discrete torsion that appears 
in the closed string Hilbert space. 

On general grounds, 
this intimate connection between discrete torsion  (in the closed  string picture) and projective representations (in the open string  picture)
has a close connection to SPT phases, which are classified by projective representations (the second group cohomology).   
In the above,
we have identified  analogues of discrete torsion phases within BCFTs and in terms of boundary states
to make this connection.
In string theory, however, it seems uncommon to assign different discrete torsions to different (fractional branes localized at) 
orbifold fixed points -- if one fixes a discrete torsion once for all for the theory, one needs to use the same discrete torsion all times. 
In the physics of SPT phases, however, we consider
different discrete torsions for different boundary states at different orbifolds fixed points.

In identifying
$\varepsilon(g|h)$ in  Eq.\ \eqref{discrete torsion and two-cocyles} as a discrete torsion,
the symmetry-enforced vanishing of the projective characters (see descriptions around \eqref{useful id}) plays an important role.
It was argued that 
the relation between discrete torsions and two-cocyles 
can be inferred by factorising the  cylinder amplitude between two fractional branes
in the closed string channel. 
Since the factorization in the closed string channel will be achieved by constructing boundary states for 
the D-branes with discrete torsion,
this consistency check amounts to 
verifying  that these  boundary states are well-projected,
and 
to checking that 
from $\mathcal{H}_h$ 
only states invariant under the $N_h$ projection contribute to the amplitude. 
This can be checked by using 
the symmetry-protected vanishing of the projective characters,
$\rho(h)=0$,
when there is a $g$ which commutes with $h$ and 
$\varepsilon(g|h)\neq 1$.

\bibliography{refKen6}
\end{document}